\documentclass[aps,twocolumn,prb,superscriptaddress,noshowpacs]{revtex4-2}

\usepackage{graphicx,nicefrac}
\usepackage{amssymb}
\usepackage{amsfonts}
\usepackage{xcolor}
\usepackage{amsmath}
\usepackage{siunitx}
\usepackage{soul}
\usepackage{flushend}

\usepackage[utf8]{inputenc}

\usepackage{hyperref}

\usepackage{amsmath}
\usepackage{mathtools}
\usepackage{dsfont}
\usepackage{bigints}
\usepackage{graphicx}
\usepackage{amssymb}
\usepackage{xcolor}
\hypersetup{
	colorlinks,
	linkcolor={red!95!black},
	citecolor={green!60!black},
	urlcolor={blue!95!black}
}

\usepackage{comment}
\usepackage{environ}

\newif\ifSM
 
\SMtrue

\ifSM
\NewEnviron{supplementalMaterials}{%
	\clearpage
	\BODY
}
\else
\excludecomment{supplementalMaterials}
\fi

\makeatletter
\def\maketitle{
	\@author@finish
	\title@column\titleblock@produce
	\suppressfloats[t]}
\makeatother

\usepackage{bbold}
\usepackage{soul}

\graphicspath{ {./Images/} }

\newcommand{\approxpropto}{\mathrel{\vcenter{
			\offinterlineskip\halign{\hfil$##$\cr
				\propto\cr\noalign{\kern2pt}\sim\cr\noalign{\kern-2pt}}}}}
\newcommand{\nodagger}{ 
	{\vphantom{\dagger}}
}

\newcommand{\specialcell}[2][c]{%
	\begin{tabular}[#1]{@{}c@{}}#2\end{tabular}}

\newcommand{\ket}[1]{\left| #1 \right>} 
\newcommand{\bra}[1]{\left< #1 \right|} 

    \newcommand{\threej}[6]{ \begin{pmatrix}
       #1 & #2 & #3 \\
       #4 & #5 & #6 
    \end{pmatrix}}

\begin{document}
	
	\title{Bisognano-Wichmann Hamiltonian for the entanglement spectroscopy of fractional quantum Hall states}
	\author{A. Nardin}
	\email{alberto.nardin@universite-paris-saclay.fr}
	\affiliation{Universit\'e Paris-Saclay, CNRS, LPTMS, 91405 Orsay, France}

	\author{R. Lopes}
	\affiliation{Laboratoire Kastler Brossel, Coll\`ege de France, CNRS, ENS-PSL University,\\ Sorbonne Universit\'e, 11 Place Marcelin Berthelot, 75005 Paris, France}
	
	\author{M. Rizzi}
	\affiliation{Institute for Theoretical Physics, University of Cologne, D-50937 K\"oln, Germany}
	\affiliation{Forschungszentrum Jülich GmbH, Institute of Quantum Control,\\
	Peter Gr\"unberg Institut (PGI-8), 52425 J\"ulich, Germany}

	\author{L. Mazza}
	\affiliation{Universit\'e Paris-Saclay, CNRS, LPTMS, 91405 Orsay, France}
	
	\author{S. Nascimbene}
	\affiliation{Laboratoire Kastler Brossel, Coll\`ege de France, CNRS, ENS-PSL University,\\ Sorbonne Universit\'e, 11 Place Marcelin Berthelot, 75005 Paris, France}
	
	\begin{abstract}
		We study the Bisognano-Wichmann Hamiltonian for fractional quantum Hall states defined on a sphere and explore its relationship with the entanglement Hamiltonian associated to the state.
		We present results for several examples, namely the bosonic Laughlin state stabilized by contact two-body interactions and the bosonic Moore-Read state stabilized by either three- or two-body interactions.
		Our findings demonstrate that the Bisognano-Wichmann Hamiltonian provides a reliable approximation of the entanglement Hamiltonian as a fully-local operator  that can be written without any prior knowledge of the specific state under consideration.
	\end{abstract} 
	
	\maketitle
	
	\section{Introduction}
	
	Exotic phases of matter exhibiting a fractional quantum Hall (FQH) effect evade the standard paradigm of symmetry breaking~\cite{wen_colloquium_2017}. 
	They can instead be classified according to a topological order encoded in the ground state's patterns of long-range entanglement~\cite{Chen_PRB_2010}.
	Various signatures of topological order have been identified, including  a robust ground-state degeneracy~\cite{Wen_IntJModPhysB_1990}, the existence of fractionalized excitations~\cite{Laughlin_PRL_1983, Arovas_PRL_1984, Nardin_PRB_2023}, the presence of chiral edge-modes~\cite{Wen_PRB_1991, Chang_RMP_2003}, and a quantized topological entanglement entropy~\cite{levin_detecting_2006,kitaev_topological_2006}. 
	
	In order to gain a deeper understanding of the structure of a FQH state, Li and Haldane~\cite{Li_PRL_2008} emphasized the importance of
	partitioning the space into two complementary regions, $A$ and $B \equiv \bar A$.
	Given the many body state $\ket{\Psi}$ defined on the entire space, they suggested to take the reduced density matrix $\rho_A = \text{Tr}_B{\ket{\Psi}\hspace{-0.11cm}\bra{\Psi}}$ obtained after tracing out any degree of freedom that does not belong to the spatial region $A$.
	Since $\rho_A$ is a positive and Hermitian operator,
 	with eigenvalues comprised between $0$ and $1$, 
	it is always legitimate to interpret it as a thermal state of the form $\rho_A \sim e^{- H_A}$. $H_A$ is the \textit{entanglement Hamiltonian} (EH), it is  dimensionless and formally defined as
 	\begin{equation}
 		\label{eq:entanglementHamiltonian}
		{H}_A=-\log{\rho}_A.
 	\end{equation}
	The spectrum of $H_A$ is called the \textit{real-space entanglement spectrum} (RSES), and it was conjectured to exhibit the same structure as a topological edge mode, according to the very suggestive and physical idea that the spatial partition cut acts as a sharp edge in the entanglement Hamiltonian~\cite{Li_PRL_2008}.
	This property can also be used as a tool to deduce the structure of the edge excitations from the ground-state wavefunction only~\cite{regnault_entanglement_2017}.
	
	Since measuring the entanglement spectrum presents a notable experimental challenge~\cite{dalmonte_entanglement_2022, choo_measurement_2018, Sankar_PRL_2023, pichler_measurement_2016,johri_entanglement_2017,beverland_spectrum_2018,kokail_entanglement_2021, joshi_exploring_2023-2}, an interesting alternative consists in considering the EH as a physical Hamiltonian defined in subregion $A$ and studying it by making a quantum simulation. 
	This physical realization of the EH maps the entanglement spectrum
    to an energy spectrum, up to a multiplicative energy scale, which is more easily accessible for experimental measurement~\cite{dalmonte_quantum_2018}.
	This approach relies on the Bisognano-Wichmann (BW) theorem~\cite{bisognano_duality_1975,bisognano_duality_1976}, which provides an exact form for $ H_A$ in terms of a local deformation of the original Hamiltonian $ H$ in the context of high-energy physics, with local and Lorentz-invariant quantum field theories. 
	
    The BW theorem is not generically exact in condensed-matter physics and in recent years many Hamiltonians inspired by the BW theorem have been proposed that provide a good approximation of the actual EH of the model, $H_A$~\cite{giudici_entanglement_2018,mendes-santos_entanglement_2019,zhang_lattice_2020, kokail_quantum_2021, zache_entanglement_2022}. 
	In order to understand how these approximations are obtained,
	it is useful to get back to the original high-energy-physics theorem, that states that,
	once we introduce the Hamiltonian density $\mathcal H (\mathbf{r})$, such that $ H = \int \mathcal H(\mathbf r) d \mathbf r$,  the exact form of $ H_A$ reads 
	\begin{equation}
		\label{eq:BWTheorem}
		 H^{\rm BW}_A = 
		\int_{\mathbf{r}\in A} \beta(\mathbf{r}) \mathcal{H}(\mathbf{r}) d\mathbf{r}.
	\end{equation}
	$\beta(\mathbf{r})$ is a deformation factor, that is a function of the distance from the partition cut. In the case of a partition of two-dimensional space where region $A$ is the half-plane $x \geq 0$, $\beta(\mathbf r)$ is proportional to $x$.

	In the condensed-matter setting, this form has been easily adapted to quantum lattice models: 
    the couplings between neighboring sites become space-dependent and are deformed according to the function $\beta(\mathbf r)$.
    When dealing with square lattices, they are deformed by a factor whose strength is linear in the distance from the bipartition cut.
    The results reported so far have underlined that whenever the low-energy description of the lattice model is described by a quantum field theory that displays Lorentz-invariance, the BW theorem allows to construct a BW Hamiltonian that is a good qualitative and quantitative approximation of the exact entanglement Hamiltonian~\cite{giudici_entanglement_2018,mendes-santos_entanglement_2019,zhang_lattice_2020, kokail_quantum_2021, zache_entanglement_2022}.
    Note that the BW Hamiltonian has been used both for topological and non-topological systems.

	Despite these technical difficulties, the BW Hamiltonian has led to the physical realization of the EH of a non-interacting quantum Hall system with ultracold atomic gases using a synthetic dimension encoded in the internal spin of an atom; in this context, the spatial deformation $\beta(\mathbf{r})$ of the BW Hamiltonian has been engineered through the inhomogeneity of spin couplings~\cite{redon_realizing_2024}.
	Yet, the use of the BW Hamiltonian to characterize the EH of FQH state in conjunction with the Li-Haldane conjecture has not been addressed so far.

	In this article, we study the link between the entanglement Hamiltonian and the BW Hamiltonian for strongly-interacting FQH states. 
	For numerical convenience, we consider the spherical geometry.
    Differently from earlier studies on BW Hamiltonian, that focused on lattice problems, we consider a system in the continuum and need to associate
 	the distance from the line cut to a quantum mechanical operator; we propose to construct the BW Hamiltonian by taking the anticommutator between $\beta(\mathbf r)$ and $\mathcal H (\mathbf r)$.
	The result of our study is that the BW Hamiltonian provides an exact long-wavelength approximation of the entanglement Hamiltonian in situations where an analytically-known model FQH wavefunction is the exact ground state of a Hamiltonian.
	We study also situations where the ground state of a FQH problem is not analytically known: in these cases we find that the BW Hamiltonian is an excellent approximation of the exact entanglement Hamiltonian.
	One point that we want to stress is that in this work we do not only focus on spectral properties (i.e.~on entanglement eigenvalues), but we also study the entanglement eigenvectors. 
	For this reason, we argue that with the help of the BW Hamiltonian we can reconstruct the most relevant part of the reduced density matrix of the many-body state.

The article is organized as follows.
In Sec.~\ref{Sec:SingleParticle} we show that the BW theorem applies to single-particle states within the lowest Landau level (LLL) and discuss the consequences of this fact on non-interacting fermionic problems related, for instance, to the integer quantum Hall effect. 
In Sec.~\ref{Sec:Model:FQH}
we then focus on systems of bosonic atoms with two- or three-body contact interactions, which are known to exhibit bosonic Laughlin and Moore-Read states. 
Through a combination of analytical and numerical calculations, we establish that the lowest-energy-branch of the BW Hamiltonian is gapped and matches the conformal-field-theory counting of a topological edge mode. 
In Sec.~\ref{Sec:FQH:NonModel} we focus on non-model wavefunctions by studying a Moore-Read state stabilized by two-body contact interactions and show that the BW Hamiltonian is also a good approximation of the entanglement Hamiltonian.
We conclude in Sec.~\ref{Sec:Conc} that for model wavefunctions, the BW Hamiltonian and the EH share the same eigenvectors, whereas in realistic cases this is true with a high level of accuracy.    
Overall, our findings provide valuable insights on the real-space entanglement Hamiltonian of FQH systems.

\section {Single-particle BW Hamiltonian and non-interacting fermionic systems}
\label{Sec:SingleParticle}

\subsection{Single-particle BW Hamiltonian}

Before delving into the FQH interacting problem and the Li-Haldane conjecture, we first focus on the real-space EH for the single-particle problem.
To mitigate any edge effects, we conduct our investigations using a spherical geometry.  
Note that the sphere can be viewed as a compactified plane, and a stereographic projection would allow us to map our results to the planar geometry.

We examine the dynamics of a particle of charge $q$ and mass $M$ as it moves on the  surface of a sphere with radius $R$. The particle is subjected to a radial magnetic field $B$ generated by a Dirac magnetic monopole of charge $Q$ positioned at the centre of the sphere. The magnetic flux through the sphere is an integer multiple $2Q\in\mathbb{N}$ of flux quanta $h/q$, so that the magnetic field is $B = \hbar Q/ (qR^2)$.
Using the latitude gauge for the vector potential $\mathbf{A}=-\hbar Q/qR\,\cot(\theta){\boldsymbol e_{\phi}} = -BR \,\cot(\theta){\boldsymbol e_{\phi} }$, where $(\theta, \phi)$ are the spherical angles, the single particle Hamiltonian reads $ H_0={ \Lambda}^2/(2M R^2)$, with $\mathbf{\Lambda}=\mathbf{R}\times\left(-i\hbar \boldsymbol{\nabla}-q \mathbf{A}\right)$~\cite{Greiter_PRB_2011,jainCompositeFermionsBook_2007}.
    
Owing to rotational symmetry, the eigenstates can be labelled by the angular momentum quantum numbers $L$ and $m$, associated to the operators $ L^2$ and $L_z$, and will be noted $\psi_{L,m}$. The spectrum organises in families of degenerate levels, the Landau levels, $E_L=\hbar\omega_c(L(L+1)-Q^2)/2Q$, where $\omega_c=qB/M$ is the cyclotron frequency, and  the angular momentum is restricted to $L=Q+n$ ($n\in\mathbb{N}$).
The LLL corresponds to $n=0$ and the single particle wave-functions $\psi_{Q,m}\propto u^{Q+m}v^{Q-m}$ are usually written in terms of spinor variables, $u(\theta, \phi)=\cos(\theta/2)e^{i\phi/2}$ and $v(\theta, \phi)=\sin(\theta/2)e^{-i\phi/2}$, with $-L \leq m\leq L$~\cite{Haldane_PRL_1985}, so that in real-space they read:
\begin{equation}
 \psi_{Q,m} \propto  
 \sin^Q(\theta)
 \cot^m  \left( \theta / 2 \right) 
e^{i m \phi}.
\end{equation}
These orbitals have probability densities centred around $\cos (\theta) = m/Q $.

\begin{figure}[t]
        \includegraphics[width=\columnwidth,trim=0 1mm 0 0]{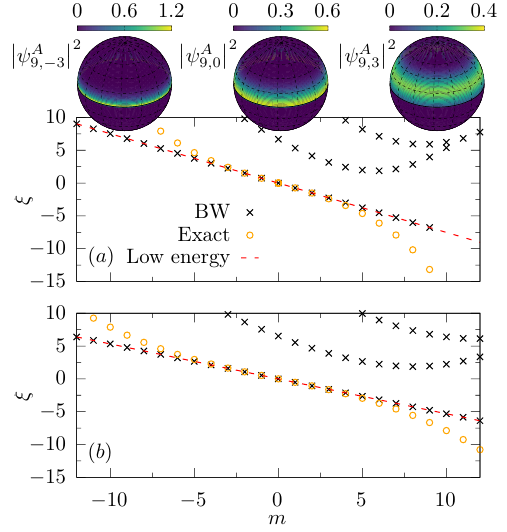}
        \caption{Single-particle RSES $\xi_A(m)$ for LLL states in Eq.~\eqref{Eq:Ent:Spectrum} [yellow circles] and spectrum of the corresponding BW Hamiltonian in Eq.~\eqref{eq:freeBWHamiltonian} [black crosses] for a bipartition at the equator and $Q=9$ (a) and $Q=18$ (b): the lowest branch of the latter spectrum corresponds to the long-wavelength linearized limit of the EH [dashed red line]. 
        The lowest branch is subject to the constraint $m\leq Q$; this restriction does not apply to higher branches.
        The three wavefunctions $\psi_{Q,m}^A$ plotted on top are defined on region $A$  and are eigenfunctions of both the EH and the BW Hamiltonian.}
        \label{fig:singleparticle}
	\end{figure}

We now consider a bipartition of the sphere into two regions $A$ and $B$, where  $A$ corresponds to the northern hemisphere $0 \leq \theta \leq \pi/2$ and $B$ to the southern one.
	This partition preserves the rotational symmetry around the $z$-axis, so that both the reduced density matrix $\rho_A = e^{- H_A}$ and the entanglement Hamiltonian $H_A$ can be decomposed over angular-momentum projections $m$. 
	
The EH of the single-particle problem  reads
	$H_A = \sum_m \xi_A(m) \ket{\psi^A_{Q,m}} \hspace{-0.13cm} \bra{\psi_{Q,m}^A}$; the eigenvalues read:
\begin{equation}
\label{Eq:Ent:Spectrum}
 \xi_A(m) = - 2 \text{atanh} \left(
 2 \frac{\text{B}_{\frac{1}{2}}(Q+1-m, Q+1+m)}{\text{B}(Q+1-m,Q+1+m)}-1
 \right),
\end{equation}
where $\text{B}(z_1, z_2)$ is the beta function and $\text{B}_x(z_1, z_2)$ is the incomplete beta function.
Like in the original LLL, the quantum number $m$ takes integer values in $-L \leq m \leq L$, the expression in~\eqref{Eq:Ent:Spectrum} being otherwise ill-defined~\cite{regnault_entanglement_2017}.
The wavefunctions $ \psi^A_{Q,m}$ are obtained by truncating at $\theta = \pi/2$ and normalising, 
\begin{equation}\psi_{Q,m}^A \propto \psi_{Q, m} \Theta_{\rm H}(\pi/2 - \theta), 
\label{Eq:Truncated:Eigenfunctions}
\end{equation}
where $\Theta_{\rm H}(x)$ is the Heaviside function. 
Details of the calculations are presented in Appendix~\ref{App:A:EntHam};
both the spectrum and the eigenfunctions are plotted in Fig.~\ref{fig:singleparticle} in some interesting cases.

Inspired by the BW theorem,
we write a BW Hamiltonian for the single-particle problem as
	\begin{equation}
		\label{eq:freeBWHamiltonian}
		 H_A^{\rm BW} =\frac{4 \sqrt Q}{\sqrt{\pi}}
		\,
		\left\{
		\frac{ H_0 - \epsilon}{\hbar\omega_c},
		\cos  \theta    \right\},
		 \quad \text{for } 0 \leq \theta \leq \frac \pi 2.
	\end{equation}
	The spatial deformation factor $\cos  \theta $, which acts as a regularised distance from the equator, is the essence of the BW approach. 
	It disconnects the regions $A$ and $B$, so that it is legitimate to consider the restriction of $H_A^{\rm BW}$ to subregion $A$. 
	The energy offset
	$\epsilon$ reflects the fact that $H_0$ is defined up to a constant.
	Contrary to the writing in Eq.~\eqref{eq:BWTheorem} that is typical of high-energy physics, where the deformation factor was a scalar, in this condensed-matter problem the $\cos \theta$ factor
	is a quantum mechanical operator, hence the requirement of the anticommutation to get an Hermitian operator. 
	In the following paragraphs, we summarize the results of the study of $H^{\rm BW}_A$ that we performed and refer the reader to Appendix~\ref{App:A} where all technical details are reported.
	Note that a similar analysis is at the basis of the experimental work reported in Ref.~[\onlinecite{redon_realizing_2024}].

	By selecting an energy offset $\epsilon=\hbar\omega_c(1+1/2Q)$, which lies midway between the lowest and first excited LL,	
    the truncated LLL wavefunctions $\psi_{Q,m}^A$ in Eq.~\eqref{Eq:Truncated:Eigenfunctions}
	are exactly the eigenfunctions of the lowest-lying energy band of the BW Hamiltonian in Eq.~\eqref{eq:freeBWHamiltonian}.
    This result seems a very solid starting point for using the BW Hamiltonian to approximate the entanglement Hamiltonian and hence in the rest of the paper we stick to it, without exploring the possibilities offered by other values.
	
	The eigenvalues of the BW Hamiltonian associated to the wavefunctions $\psi^A_{Q,m}$ read:
	\begin{equation}
		\label{eq:bisognanoEnergies}
		\xi^{\rm BW}_A (m) = -\frac{4}{\sqrt{\pi\,Q}}\,m,  
		\qquad \text{for } m \leq Q;
	\end{equation}
	this result can be checked by direct calculation.
	The full spectrum of the BW Hamiltonian is plotted in Fig.~\ref{fig:singleparticle}. The lowest band aligns with the EH exact result in the thermodynamic limit ($Q=R^2/l_B^2 \to \infty$, with $l_B = \sqrt{\hbar / (qB)}$ the magnetic length)~\cite{Rodriguez_PRB_2009,Rodriguez_PRL_2012,Sterdyniak_PRB_2012,Dubail_PRB_2012a,Dubail_PRB_2012b, Oblak_PRB_2022}, thus providing its long-wavelength limit. Indeed, an expansion at lowest order in $m$ of the expression for $\xi_A(m)$ in Eq.~\eqref{Eq:Ent:Spectrum} yields exactly Eq.~\eqref{eq:bisognanoEnergies}, see also App.~\ref{App:A:EntHam} for more details. The two panels in Fig.~\ref{fig:singleparticle} compare two values of $Q$ and show that by increasing $Q$ the interval of values of $m$ for which the BW eigenvalues are a good approximation of the $\xi_A(m)$ is larger.

	The BW Hamiltonian is here defined only in the northern hemisphere, since we want to reproduce the reduced density matrix in this region. Due to this domain restriction, the BW spectrum
	extends all the way to $m \to - \infty$ since the wavefunctions $\psi_{Q,m}\propto u^{Q+m}v^{Q-m}$ can be normalised even for $m<-Q$ once restricted to the northern hemisphere, in contrast to the restricted range for the RSES. 
	Hence, the BW Hamiltonian $H_{A}^{\rm BW}$, that is defined on the northern hemisphere, does not experience this normalization problem.
	Considering the southern hemisphere instead would lead to a different restriction, $m > -Q$.
	Interestingly, the BW Hamiltonian  offers an efficient approach to investigating infinite and precisely linear spectra on finite systems, which may prove useful in various contexts.

\subsection{
Non-interacting many-body fermionic states}\label{Sec:Non:Interacting:Fermions}

The results that we just presented for single-particle states can be easily employed to make predictions on the entanglement Hamiltonian of many-body problems with non-interacting fermions~\cite{regnault_entanglement_2017}.

We study the $N$-particle fermionic state where the LLL is completely filled; 
we consider the bipartition at the equator $\theta = \pi/2$ and focus on the northern hemisphere.
Since the bipartition is defined by a fixed value of the polar angle $\theta$, preserving rotational symmetry around $z$, the reduced density matrix $\rho_A$  (and consequently the entanglement Hamiltonian) can be decomposed into symmetry sectors characterized by a fixed particle number $N_A$  and angular momentum $ L_z^A$ (see Appendix~\ref{App:QuantumNumbersRedRho} for an explicit proof).
For each sector with well-defined $N_A$ and $L_z^A$ values, we can define an entanglement Hamiltonian; such an operator is defined on the Hilbert space spanned by the states obtained by putting $N_A$ particles on LLL orbitals such that the total angular momentum is $L_z^A$. 
A simple calculation~\cite{regnault_entanglement_2017} shows that these states are eigenvectors of the entanglement Hamiltonian with eigenvalue equal to the sum of the single-particle entanglement eigenvalues of the occupied orbitals, $\sum_{m \text{ occupied}} \xi_A(m)$, and eigenvector given by the Slater determinants of the single-particle entanglement eigenvectors $\psi_{Q,m}^A$ that are occupied.

The BW Hamiltonian that approximates this EH is the many-body version of the single-particle Hamiltonian in Eq.~\eqref{eq:freeBWHamiltonian}, generalized to the $N_A$-particle case:
\begin{equation}
 H^{\rm BW}_A = \frac{4 \sqrt{Q}}{\sqrt{\pi}}
		 \sum_{i=1}^{N_A} \left\{
		\frac{H_0(i) - \epsilon}{\hbar \omega_c}, \cos \theta_i 
		\right\}.
		\label{Eq:BW:NA:free}
\end{equation}
This BW Hamiltonian features a lowest-energy branch that reproduces the features of the exact EH.
These low-energy eigenvalues are given by $\sum_{m \text{ occupied}}\xi^{\rm BW}_A(m)$ and thus constitute a long-wavelength approximation of the exact entanglement eigenvalues.
Moreover, since $\xi^{\rm BW}_A(m)$ is exactly proportional to $m$, the many-body entanglement eigenvalue reads:
\begin{equation}
 \xi^{\rm BW}_A = - \frac{4}{\sqrt {\pi Q}} L_z^A. \label{Eq:EVL:Chiral}
 \end{equation}
The eigenvectors associated to this lowest-energy branch coincide precisely with the eigenvectors of the exact EH, a property that is directly inherited from that of the single-particle eigenvectors of the BW Hamiltonian~\eqref{eq:freeBWHamiltonian}. 

The reader should note that Hamiltonian~\eqref{Eq:BW:NA:free} features many more energy levels, obtained by considering states of the single-particle BW Hamiltonian~\eqref{eq:freeBWHamiltonian} that are not in its lowest-energy branch. These states do not have a direct counterpart in the exact EH: they should be though of as states whose entanglement eigenvalue is infinite.
Discarding these states is equivalent to the procedure that is routinely applied in the FQH context of projecting the complex many-body Hamiltonian of interacting electrons onto the LLL. 

We can perform this procedure also in the context of the BW Hamiltonian, as it gives rise to a simple second-quantisation expression that is very useful for a numerical study. 
To this goal, we introduce the projection operator $\Pi_{\rm BW}$ onto the lowest-energy band of the BW Hamiltonian and the fermionic operators $c_{m}$ associated to the wavefunctions $\psi_{Q,m}^A$, with $m\leq Q$; canonical anticommutation relations are satisfied. 
The projected BW Hamiltonian~\eqref{Eq:BW:NA:free} reads:
\begin{align}
 \Pi_{\rm BW} H_A^{\rm BW} \Pi_{\rm BW} =& \sum_{m \leq Q}
 \xi_A^{\rm BW}(m) c^\dagger_m c_m = \nonumber \\
 =& - \frac{4}{\sqrt{\pi Q}}
 \sum_{m \leq Q} m \, c^\dagger_m c_m = - \frac{4}{\sqrt{\pi Q}} L_z.
\end{align}
The latter equality is the operatorial version of Eq.~\eqref{Eq:EVL:Chiral}, which instead deals with eigenvalues.
One easily recognizes in this model the chiral Luttinger liquid model, or chiral boson model, that describes the edge of an integer quantum Hall state, showing in a first example that the BW approach is compatible with the Li-Haldane conjecture.

This analysis shows that the single-particle EH in the long-wavelength limit can be expressed as a \textit{local operator}.
Note that we are not presenting here an effective boundary Hamiltonian that mimics the spectrum of $H_A$, as it has already been discussed~\cite{Dubail_PRB_2012a,Dubail_PRB_2012b,Henderson_PRB_2021}, but a true bulk Hamiltonian defined on the entire region $A$, that approximates both the eigenvalues and the eigenvectors of $H_A$.
Prior to this study, it was not immediately apparent that such a formulation was feasible. 
One further notable feature of Eq.~\eqref{eq:freeBWHamiltonian} is that it does not require prior diagonalization of the Hamiltonian $H$ in order to express the BW Hamiltonian $H_A^{\rm BW}$.

\section{BW Hamiltonian for model FQH wavefunctions}
\label{Sec:Model:FQH}

We now study the entanglement properties of the ground state of $N$ \textit{interacting} particles  on the sphere. 
As before, we will consider a cut passing through the equator, separating the northern hemisphere from the southern one.
The reduced density matrix $\rho_A$ (and hence the EH) can be decomposed into symmetry sectors with a fixed number of particles $N_A$ and angular momentum $L_z^A$, see the discussion in Sec.~\ref{Sec:Non:Interacting:Fermions} and the detailed proof in Appendix~\ref{App:QuantumNumbersRedRho}.

In general, the study of the EH in a given sector with $N_A$ particles requires computing the ground state for a larger atom number $N > N_A$. In contrast, the direct study of the BW Hamiltonian $H_A^{\rm BW}$ only requires the calculation of the spectrum for $N_A$ particles, which is less computationally demanding.
Moreover, as we will see, also in the interacting case writing down the BW Hamiltonian does not require the prior knowledge of the ground state of the Hamiltonian $H$.
In interacting problems as those on which we are going to focus on, this constitutes a clear advantage.
In this section, however, we focus on FQH Hamiltonians whose ground state is known exactly; situations where the ground state can only be computed numerically are discussed in Sec.~\ref{Sec:FQH:NonModel}.

\subsection{Laughlin state}
    
\begin{figure}[t]
\includegraphics[width=\columnwidth
,trim=0 3mm 0 0]{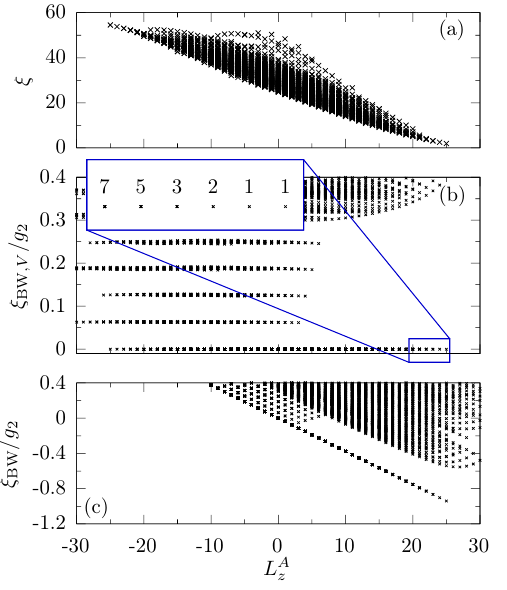}
\caption{(a) RSES of a bosonic Laughlin state of $N=10$ particles (corresponding to $Q=9$), with $N_A=5$.
(b) Spectrum of the BW Hamiltonian $\Pi_{\rm BW} V_A^{\rm BW} \Pi_{\rm BW} $ for $N_A=5$ bosons; the plot does not depend on $g_2$, which is simply a multiplicative constant of the eigenvalues.
The lowest lying zero-energy band has the same degeneracy of the RSES at the same $L_z^A$. 
The inset shows a close-up small portion (highlighted by the blue rectangle) of such a band focusing on those values of $L_z^A$ for which the degeneracy matches the thermodynamic-limit conformal-field-theory prediction; numbers with the counting of degenerate states have been added for clarity.
(c) Spectrum of the total BW Hamiltonian $\Pi_{\rm BW}(H^{\rm BW}_A+V_A^{\rm BW}) \Pi_{\rm BW}$. Due to the presence of the free term given by Eq.~\eqref{Eq:BW:free:Many:Body}, the plot depends on $g_2$, for which we take the value $20$.
		\label{fig:Laughlin_BW_ideal}}
	\end{figure}

We begin our investigation by examining the bosonic Laughlin state at filling fraction $\nu=1/2$, which is the exact and gapped ground-state of a two-body contact interaction Hamiltonian $V_2 = \sum_{i<j}V_2(\mathbf{r}_i,\mathbf{r}_j)$ with $V_2(\mathbf{r}_i,\mathbf{r}_j)=g_2\,\delta^{(2)}(\mathbf{r}_i-\mathbf{r}_j)$~\cite{Wilkin_PRL_1998,Cooper_PRL_2001,Cooper_AdvPhys_2008};
on the sphere, its stabilization requires $Q=N-1$~\cite{Regnault_PRB_2004}.
Its RSES, shown in Fig.~\ref{fig:Laughlin_BW_ideal}(a), can be characterized by the quantum numbers $N_A$ and $L_{z}^A$. 
The ground state is at angular momentum 
$L_z^A = N_A (Q+1-N_A)$.
The entanglement spectrum consists of a finite branch of states whose counting in the thermodynamic limit corresponds to that of a chiral boson, in agreement with the Li-Haldane conjecture~\cite{Haldane_PRL_2008,Chandran_PRB_2011,Qi_PRL_2012,Swingle_PRB_2012,Dubail_PRB_2012a,Dubail_PRB_2012b, Sterdyniak_PRB_2012}.

The generalization of the BW ansatz of Eq.~\eqref{eq:freeBWHamiltonian} to the many-body case with  contact  interactions involves both the single-particle Hamiltonian $H^{\rm BW}_A$ in Eq.~\eqref{Eq:BW:NA:free} and an interaction term, for which we propose
to proceed in the following way.
Once the contact interaction on the sphere is rewritten in spherical coordinates as:
\begin{equation}
 V_2 = \sum_{i<j} g_2
 \frac{\delta (\cos \theta_i - \cos \theta_j) \delta (\phi_i - \phi_j)}{R^2},
\end{equation}
the Bisognano-Wichmann interaction $V^{\rm BW}_A$ should be obtained by anticommuting it with the $\cos \theta_i$. This is in principle problematic as the interaction term, being a two-body Hamiltonian, couples at the same time both $\cos \theta_i$ and $\cos \theta_j$ of the interacting particles labelled by $i$ and $j$; it is not obvious which one of the two (or which function of both) should be taken.
However, since we are dealing with a contact interaction, when the particles repel each other the values of $\cos \theta_i$ and $\cos \theta_j$ coincide: we can simply choose one of the two. Following this reasoning, we propose the following form for the BW two-body interaction:
\begin{equation}
    \label{eq:manyBodyBWHamiltonian}
		V_A^{\rm BW} = 
		\frac{4 \sqrt{Q}}{\sqrt{\pi}}\sum_{i<j}^{N_A}
		\left\{
		\frac{g_2 \, \delta(\cos \theta_i - \cos \theta_j) \delta(\phi_i-\phi_j) }{R^2 \hbar\omega_c},
		\cos \theta_i \right\} .
\end{equation}
	
In order to determine the low-energy features of $H_A^{\rm BW}+V_A^{\rm BW}$, we proceed in full analogy with the standard treatment of the FQH effect, based on pseudopotentials; we thus consider the projector $\Pi_{\rm BW}$ introduced in Sec.~\ref{Sec:Non:Interacting:Fermions} and study the properties of $\Pi_{\rm BW} H_{A}^{\rm BW} \Pi_{\rm BW} + \Pi_{\rm BW} V_{A}^{\rm BW} \Pi_{\rm BW}$.
    
We introduce the bosonic operators $a_m$ associated to the BW single-particle states $\psi_{Q,m}^A$, with $m \leq Q$; they satisfy canonical anti-commutation relations.
The non-interacting part of the Hamiltonian reads:
    \begin{equation} 
    \label{Eq:BW:free:Many:Body}
     \Pi_{\rm BW} H_{A}^{\rm BW} \Pi_{\rm BW} = - \frac{4}{\sqrt{\pi Q}} \sum_{m \leq Q} m a^\dagger_m a_m = - \frac{4}{\sqrt{\pi Q}}L_z.
    \end{equation}
    Note that the relation between the projected free Hamiltonian and the total angular momentum is correct both in the fermionic case discussed in Sec.~\ref{Sec:Non:Interacting:Fermions} and in the bosonic case that we discuss here as it follows only from the fact that $\xi^{\rm BW}_A(m) \sim m$.
    
    In general, the interacting component $ \Pi_{\rm BW} V_A^{\rm BW} \Pi_{\rm BW}$ is a quartic Hamiltonian that conserves both the total number of particles and the total angular momentum:
\begin{equation}
 \Pi_{\rm BW} V_A^{\rm BW} \Pi_{\rm BW} = \frac{1}{2!} \sum_{\substack{m_1,m_2\\m_3,m_4}} 
 V_{\substack{m_1,m_2\\m_3,m_4}} \, a^\dagger_{m_1} a^\dagger_{m_2} a_{m_3} a_{m_4}.
\end{equation}
Further details and the explicit form of $V_{\substack{m_1,m_2\\m_3,m_4}}$ are given Appendix~\ref{Appendix:Diag:ManyBody}; in the  rest of this section we solely focus on the numerical and analytical results regarding the low-energy properties of the projected BW Hamiltonian.
    
    \subsubsection{Eigenvalues}
    \label{Sec:Laugh:EGVL:Counting}
    
We show the results of  the numerical diagonalisation of $\Pi_{\rm BW} V_A^{\rm BW} \Pi_{\rm BW}$ in Fig.~\ref{fig:Laughlin_BW_ideal}(b) for $N_A=5$. 
A branch of zero-energy states  extends up to $L_z^A = N_A (Q+1-N_A)$, where a single state is seen; this behaviour is exactly like the one of the RSES in Fig.~\ref{fig:Laughlin_BW_ideal}(a). 
As we move towards smaller values of $L_z^A$, the degeneracy of the states increases, precisely matching the expected counting for a chiral-boson conformal field theory up to finite size effects~\cite{Wen_AdvPhys_1995}, see inset in Fig.~\ref{fig:Laughlin_BW_ideal}(b).
		We provide some additional technical information on the edge-state counting in Appendix~\ref{App:StateCounting:C}.
For completeness, in Fig.~\ref{fig:Laughlin_BW_ideal}(c) we plot the spectrum of the total BW Hamiltonian $\Pi_{\rm BW}(H^{\rm BW}_A+V_A^{\rm BW}) \Pi_{\rm BW}$, which is just a shifted version of the data in Fig.~\ref{fig:Laughlin_BW_ideal}(b) by the amount given in Eq.~\eqref{Eq:BW:free:Many:Body}.

Note that even if in Fig.~\ref{fig:Laughlin_BW_ideal}(a) the entanglement eigenvalues are not degenerate, they correspond to the degenerate zero-energy levels in Fig.~\ref{fig:Laughlin_BW_ideal}(b).
The fact that BW eigenvalues have a flat zero-energy, whereas the exact entanglement eigenvalues have a non-zero slope, is due to the fact that in the calculation of the BW spectrum we disregarded the single-particle term $\Pi_{\rm BW} H_A^{\rm BW} \Pi_{\rm BW}$ proportional to $L_z^A$.
States that appear in Fig.~\ref{fig:Laughlin_BW_ideal}(b) at energy larger than zero have to be considered as spurious effects: they appear at finite energy in the BW problem but in the exact entanglement Hamiltonian are pushed up to infinite energy (and thus do not appear in the entanglement spectrum).
	
We can conclude that the interacting BW Hamiltonian exhibits a spectrum with a low-energy branch that satisfies the Li-Haldane conjecture, which proves the interest of performing a quantum simulation of this Hamiltonian.
		
\subsubsection{Eigenvectors}
	 
To further support the relevance of the BW Hamiltonian, we provide exact analytical expressions for the zero-energy eigenstates of $\Pi_{\rm BW} V_A^{\rm BW} \Pi_{\rm BW}$, paralleling the well-known construction of excitations in a Laughlin fluid using totally-symmetric polynomials \cite{Wen_AdvPhys_1995}.
Our approach is deeply rooted in the fact that the single-particle levels are restricted to the lowest Bisognano-Wichmann band. 	
We require that the many-body wavefunction has the good exchange symmetry for a system of $N_A$ bosons, and that each particle sees a zero of the wavefunction at the position of the other ones;
in this way the many-body states will lie in the kernel of the BW interaction Hamiltonian, which can still be interpreted as the sum of pair-wise contact potentials, as highlighted by the writing in the second line of Eq.~\eqref{eq:manyBodyBWHamiltonian}.

In order to write the ground state of the BW Hamiltonian at angular momentum $L_z^A = N_A (Q+1-N_A)$, we stereographically project the sphere onto the complex plane following the standard prescription $z_i=2R\,\cot(\theta_i/2)e^{i\phi_i}=2R\,u_i/v_i$~\cite{Read_PRB_1996} and write:
\begin{equation}
 \Psi_0(\{ z_i \}) \sim \prod_{i<j}(z_i-z_j)^2\,\prod_i\, \frac{ \Theta_{\rm H}\left(
 {\frac{|z_i|}{2R} - 1}
 \right)}{\left(1+ \frac{|z_i|^2}{4R^2}\right)^{1+Q}}.
\end{equation}
This wavefunction corresponds to a truncated Laughlin state; indeed the Heaviside function comes from the restriction of the $N_A$ particles to the northern spherical cap.
Note that the probability of two particles occupying the same position is zero; the wavefunction has a zero of second order whenever $z_i$ approaches $z_j$.

We can furthermore mimic the well-known construction of excitations in a Laughlin fluid~\cite{Wen_AdvPhys_1995}, by writing
\begin{equation}
	\Psi(\{ z_i \}) \sim \Phi_{\ell} (\{ z_i \})\,\times \,\Psi_0(\{ z_i \}),
	\label{Eq:Excitations:BW}
\end{equation}
 where
$\Phi_{ \ell} (\{ z_i \})$ is a symmetric polynomial of homogeneous degree $\ell$.	
In each angular momentum sector $L_z^A$, the counting of these polynomials exactly matches that of the real-space entanglement spectrum at fixed $N_A$ and $L_z^A$, and in the scaling limit approaches the counting of the eigenstates of the chiral-boson model.
Moreover, the probability of two particles occupying the same position is zero for all these wavefunctions and thus we are guaranteed that they are in the zero-energy band of the BW interacting Hamiltonian $V_{\rm BW}^A$.
These results on the use of symmetric polynomial for writing the edge excitations of the Laughlin state have been widely established in the literature~\cite{Wen_AdvPhys_1995} and here they are simply reinterpreted in the BW context.

\begin{figure}[t]
	\includegraphics[width=\columnwidth,trim=0 1mm 0 0]{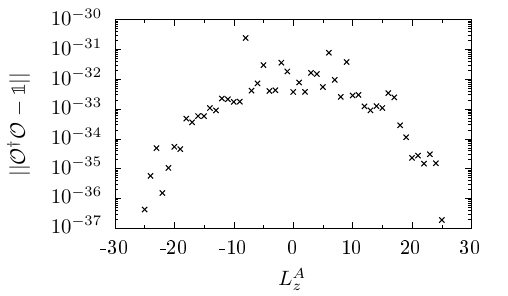}
	\caption{Deviations from unitarity of the matrix of overlaps between real-space entanglement Hamiltonian eigenvectors with finite entanglement energy and the Bisognano-Wichmann Hamiltonian eigenvectors lying below the entanglement gap. 
		The image on the left is for the case of a bosonic Laughlin state with $N=10$ particles ($Q=9$), bipartitioned at the equator in two equal hemispheres; $N_A=5$.
		}
	\label{fig:overlaps_ideal}
\end{figure}

\subsubsection{Overlap matrix}

It is important to stress that 
the states in Eq.~\eqref{Eq:Excitations:BW} exactly match the eigenvectors of the real-space entanglement Hamiltonian. 
This point can be proven analytically, but here we rather rely on the numerical verification that the vector-space spanned by the eigenvectors~\eqref{Eq:Excitations:BW} at fixed $(N_A, L_z^A)$ spans exactly the same vector-space spanned by the eigenvectors of the entanglement Hamiltonian.
We do so by computing the matrix of overlaps $\mathcal{O}_{\alpha,i}=\left<\phi^{\rm RSES}_\alpha|\psi^{\rm BW}_i\right>$  between the
eigenvectors $\ket{\psi^{\rm BW}_i}$ of the BW Hamiltonian with those of the entanglement one, $\ket{\phi^{\rm RSES}_\alpha}$; 
the calculation is performed for fixed values of $N_A$ and $L_z^A$. 

We already discussed in Sec.~\ref{Sec:Laugh:EGVL:Counting} that, at least for a certain range of values of $L_z^A$, the counting of states below the so-called ``entanglement-gap" matches for the two approaches.
Since the former case is known to satisfy the Li-Haldane conjecture, modulo finite system-size corrections, we conclude that also the BW Hamiltonian does.
As a consequence, the overlap matrix $\mathcal{O}$ restricted to such a subspace is a square matrix, which we expect to be unitary if the BW Hamiltonian has the same eigenvectors as the entanglement one. 
To test the unitarity we compute the (normalized) Hilbert-Schmidt norm of $\delta=\mathcal{O}^\dagger \mathcal{O}^\nodagger-\mathbb{1}$
\begin{equation}
	||\delta||^{2} = \frac{1}{d}\,\sum_{i,j} |\delta_{i,j}|^2
\end{equation}
where the sum runs over the $d^2$ coefficients of the difference matrix $\delta$, namely the Li-Haldane counting of the low-entanglement-energy modes. Normalizing by $d$ ensures that $||\mathbb{1}||=1$, i.e. that the identity has unit norm. 
Moreover, it accounts for finite-size effects in the Li-Haldane counting: at fixed $N_A$ and $L_z^A$, the conformal field theory counting is recovered only when $N\rightarrow\infty$.

The results are shown in Fig.~\ref{fig:overlaps_ideal}. 
It can be seen that, to numerical accuracy, the two set of eigenvectors indeed span the same vector-subspace.
We stress that, for the numerical calculation,  the single-particle spectrum, which in principle extends to $m \to -\infty$, is truncated to a finite interval $-Q\leq m\leq Q$. Although the excluded single-particle states could potentially contribute additional zero-energy states to the many-body spectrum, the truncation has no effect on the region $L_z^A >(N_A-2)(Q+1-N_A)$. Within this interval, our results demonstrate that the subspace spanned by the zero-energy branch of $H_A^{\rm BW}$ coincides with the one spanned by the eigenvectors of the EH.

	\subsection{Contact two-body interaction as a proxy for the Bisognano-Wichmann Hamiltonian}
	
\begin{figure}[t]
\includegraphics[width=\columnwidth,trim=0 1mm 0 0]{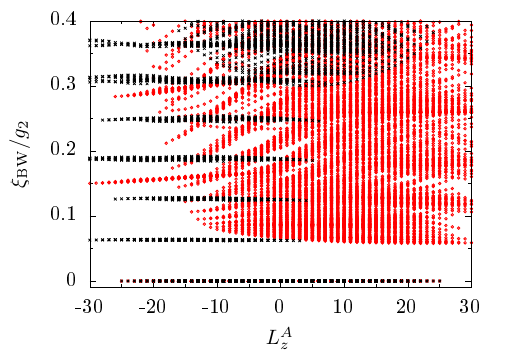}
\caption{Spectra of the pure contact interaction  $\Pi_{\rm BW} V_2 \Pi_{\rm BW}$ (red circles) and of the Bisognano-Wichmann interaction $\Pi_{\rm BW} V^{\rm BW}_A \Pi_{\rm BW}$ (black crosses) projected onto the lowest branch of the single-particle Bisognano-Wichmann Hamiltonian. We consider $N=10$, $Q=9$ and $N_1 = 5$. Both spectra feature an exact zero-energy branch with the same eigenvectors.}
\label{fig:SM_Laughlin_MR_noBW}
\end{figure}

The writing in the second line of Eq.~\eqref{eq:manyBodyBWHamiltonian} highlights the fact that $V_A^{\rm BW}$ can be interpreted as an interaction with a interaction with spatially-dependent strength $g_2 (\cos \theta_i + \cos \theta_j)/2$;
in practice, the experimental implementation of this spatially-deformed interactions  may be challenging.
From the previous section it is clear that, since the relevant many-body states of the BW Hamiltonian have zero interaction energy (we deal with contact interaction and the probability that two particles are at the same position is strictly zero), the spatial deformation associated to $\Pi_{\rm BW} V_A^{\rm BW} \Pi_{\rm BW}$ is not necessary when dealing with model wavefunctions.
This simple observation is highly relevant for quantum-simulation platforms, where in practice the experimental implementation of
spatially-deformed interactions is not straightforward: 
provided the deformed single particle BW can be implemented (as done by two of us in the experiment reported in Ref.~[\onlinecite{redon_realizing_2024}]), the ``bare" two-body interactions will stabilize states which correspond to the eigenstates of the entanglement Hamiltonian of a Laughlin state (this is true also when we consider ``bare'' three-body interactions and the Moore-Read state, as discussed in Sec.~\ref{Sec:Sub:MR}).

We have numerically verified this by examining the case of the pure contact interactions $V_2$, projected onto the BW single-particle orbitals: $\Pi_{\rm BW} V_2 \Pi_{\rm BW}$ (see Appendix~\ref{Appendix:Diag:ManyBody} for more details on this procedure); 
our result is illustrated in Fig.~\ref{fig:SM_Laughlin_MR_noBW}. 
Even though the excited states (that have non-zero energy) in the two cases are different, the zero-energy branch shares exactly the same eigenvectors (and therefore it exhibits the same state counting).
We notice that the BW Hamiltonian $\Pi_{\rm BW} V_A^{\rm BW} \Pi_{\rm BW}$  features, at angular momentum $L_z^A$ close to that of the Laughlin state, $L_z^A=N_A(Q+1-N_A)$, a larger entanglement gap when compared to $ \Pi_{\rm BW} V_2 \Pi_{\rm BW}$. We did not  further study the properties of the excited states of the two Hamiltonians.

\subsection{Moore-Read state}
\label{Sec:Sub:MR}

\begin{figure}[t]
\includegraphics[width=\columnwidth,trim=0 3mm 0 0]{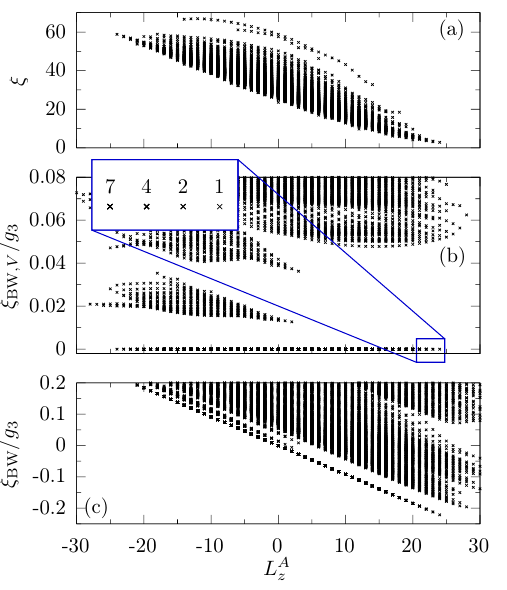}
\caption{
(a) RSES of a bosonic Moore-Read state of $N=14$ particles (corresponding to $Q=6$), with $N_A=7$.
(b) Spectrum of the projected BW Hamiltonian $V$ for $N_A=7$ bosons, without the linear dispersion term (Eq.~\eqref{eq:freeBWHamiltonian}), in the case of three-body contact interactions $V_3$; the plot does not depend on $g_3$, which is simply a multiplicative constant of the eigenvalues.
The lowest lying zero-energy band has the same degeneracy of the RSES at the same $L_z^A$. 
The inset shows a close-up small portion (highlighted by the blue rectangle) of such a band focusing on those values of $L_z^A$ for which the degeneracy matches the thermodynamic-limit conformal-field-theory prediction; numbers with the counting of degenerate states have been added for clarity.		
(c) Spectrum of the total BW Hamiltonian, including the interacting and non-interacting terms. Due to the presence of the free terms given by Eq.~\eqref{Eq:BW:NA:free}, the plot depends on $g_3$, for which we take the value $100$.
\label{fig:MooreRead_BW_ideal}}
\end{figure}

To conclude this section on model FQH wavefunctions, we consider the bosonic Moore-Read state at filling $\nu=1$ as a second case study. This state corresponds to the exact ground state of a three-body contact interaction $V_3 = \sum_{i<j<k} V_3(r_i, r_j, r_k)$, with $V_3(\mathbf{r}_i,\mathbf{r}_j,\mathbf{r}_k)=g_3\,\delta^{(2)}(\mathbf{r}_i-\mathbf{r}_j)\delta^{(2)}(\mathbf{r}_i-\mathbf{r}_k)$ \cite{Greiter_PRL_1991,Read_PRB_1999,Cappelli_NPB_2001}. 
On the sphere, its stabilization requires $2Q=N-2$~\cite{Regnault_PRB_2004}.
The extension of the BW Hamiltonian to the contact three-body interaction is straightforward; 
details on the numerical diagonalization are in Appendix~\ref{Appendix:Diag:ManyBody}.
In this case as well, the interacting BW Hamiltonian exhibits a zero-energy band that shares the same eigenvector structure as the exact entanglement Hamiltonian (see Fig.~\ref{fig:MooreRead_BW_ideal}). 
Note that, analogously to the Laughlin case analyzed above, even if in Fig.~\ref{fig:MooreRead_BW_ideal}(a) the entanglement eigenvalues are not degenerate, they correspond to the degenerate zero-energy level in Fig.~\ref{fig:MooreRead_BW_ideal}(b).
In agreement with the Li-Haldane conjecture, the state degeneracies match the edge conformal field theory, which consists for the Moore-Read state of a chiral boson and a chiral fermion (see Appendix~\ref{App:StateCounting:C} for more details on this point). 
For completeness, in Fig.~\ref{fig:MooreRead_BW_ideal}(c) we plot the spectrum of the total BW Hamiltonian, comprising both non-interacting and interacting terms.

\begin{figure}[t]
 \includegraphics[width=\columnwidth]{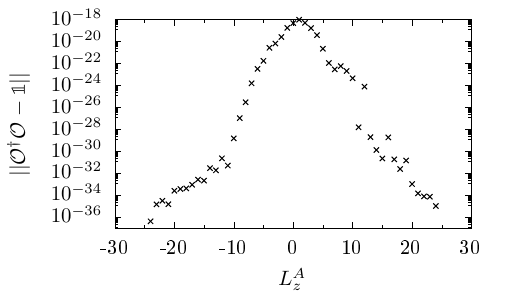}
 \caption{Deviations from unitarity of the matrix of overlaps between real-space entanglement Hamiltonian eigenvectors with finite entanglement energy and the Bisognano-Wichmann Hamiltonian eigenvectors lying below the entanglement gap. 
            The plot is for the case of a bosonic Moore-Read state with $N=14$ particles ($Q=6$), bipartitioned at the equator in two equal hemispheres; $N_A=7$.}
 \label{Fig:Overlap:MR}
\end{figure}

We now parallel the analysis that we have presented before and discuss the eigenvectors of the BW Hamiltonian.
Fig.~\ref{Fig:Overlap:MR} presents the comparison of the eigenvectors of the BW Hamiltonian and of the entanglement Hamiltonian; 
the comparison is excellent and it shows that also in this case the BW Hamiltonian has the same eigenvectors as $H_A$ because the overlap matrix $\mathcal O$ is unitary up to numerical accuracy.

	\section{BW Hamiltonian for non-model FQH states.}
	\label{Sec:FQH:NonModel}
	
	The exactness of the results presented so far for 
	analytically-known 
	model wavefunctions opens up the question on whether the BW Hamiltonian can be used to study the RSES of generic FQH states.
	In this final section we present a positive answer 
	by considering the state in the Moore-Read phase stabilized by two-body contact interactions~\cite{Regnault_PRL_2003,Regnault_PRB_2004,Regnault_PRA_2005,Cooper_AdvPhys_2008,Thomale_PRL_2010}, which is a state relevant for cold-atom experiments~\cite{palm_bosonic_2021} and whose wavefunction cannot be expressed analytically. 
We first present numerical data comparing the entanglement eigenvalues of this state with those of the BW Hamiltonian.
We then show that the eigenvectors of the BW Hamiltonian are, to a good level of approximation, the same as those of the EH, thereby proving the usefulness of the BW ansatz.

\subsection{Eigenvalues}
	
		\begin{figure}[t]
		\includegraphics[width=\columnwidth,trim=0 2mm 0 0]{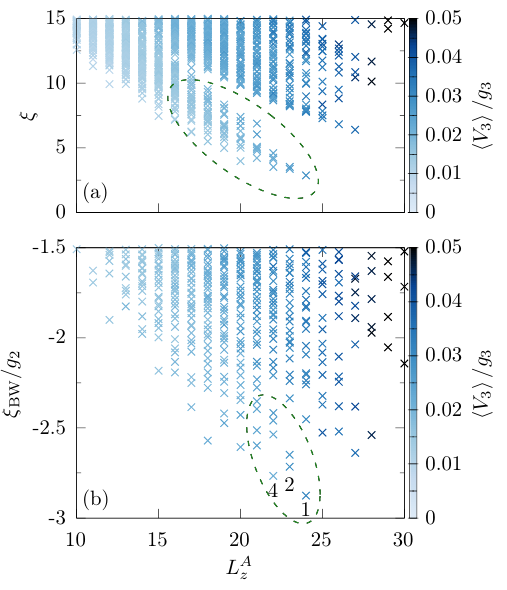}
		\caption{
			(a) RSES ($N_A=7$) of a state of $N=14$ bosons, $Q=6$, stabilized by two-body contact interactions $V_2$, which is expected to be in the Moore-Read universality class.
			(b) Spectrum of the BW Hamiltonian $\Pi_{\rm BW} V_A^{\rm BW} \Pi_{\rm BW}$ for $N_A=7$ bosons. In both cases the points have been coloured according to the expectation value of the three-body contact interaction $V_3$.		
			For clarity reasons an arbitrary linear dispersion term has been introduced in the BW case.
            For these data, $g_2 = \frac{4}{0.2 \sqrt{\pi Q}} \simeq 4.61$.
            \label{fig:MooreRead_BW_nonIdeal}}		
	\end{figure}

	\begin{figure*}[t]
\includegraphics[width=\linewidth,trim=0 1mm 0 0]{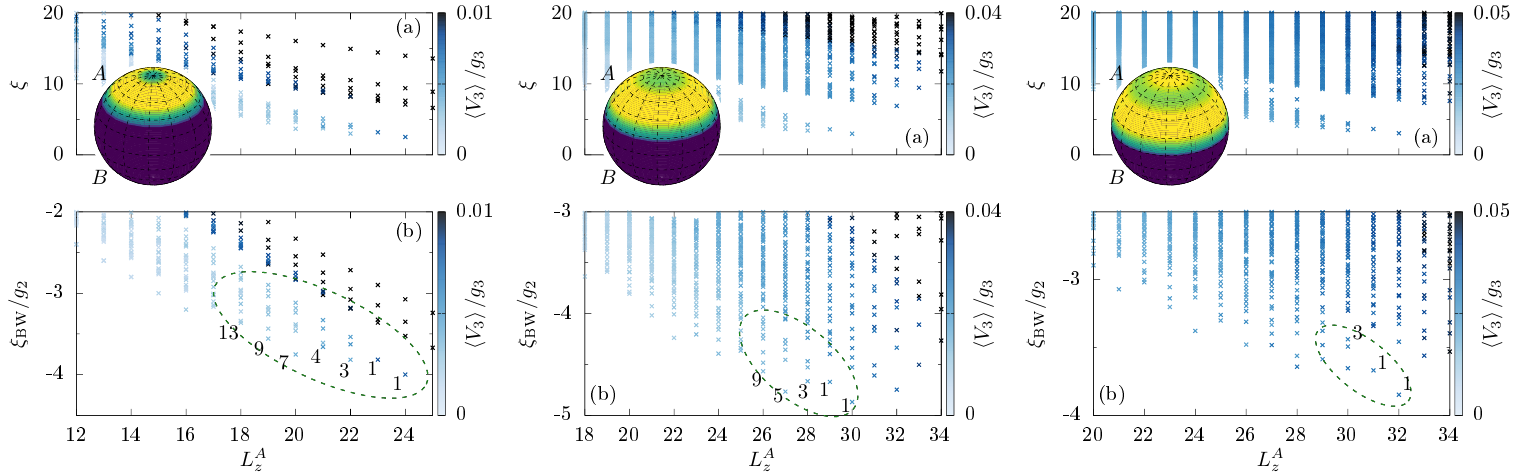}
	\caption{The three images show the real-space entanglement spectra for $N=16$, $Q=7$ bosonic Moore-Read state at filling $\nu=1$, stabilized by two-body contact interactions (panels (a)), compared to the Bisognano-Wichmann ansatz for the given bipartiton (panels (b)), which, from left to right, are given by $N_A=4,6,8$. 
	In the lower panel, the dashed elliptical shape serves as a guide to the eye for the lower branch of states and the numbers indicate the number of states at a given $L_z^A$. These states should be compared with the lower branch that appears in the upper panels with the RSES.
		The insets show the density of the lowest-lying state of the Bisognano-Wichmann Hamiltonian as a heat-map, illustrating how the spherical cap at longitude $\theta_0$ is varied as a function of $N_A$. 
		In all cases the points have been coloured according to the expectation value of the three-body contact interaction $V_3$. For clarity reasons an arbitrary linear dispersion term has been introduced in the Bisognano-Wichmann case, so that the comparison with the plots in the upper row is facilitated.
		For these data, $g_2 = \frac{4}{0.2 \sqrt{\pi Q}} \simeq 4.26$.}
	
	\label{fig:SM_cuts}
\end{figure*}
	
In previous works, it has been shown that the RSES of this state, plotted in Fig.~\ref{fig:MooreRead_BW_nonIdeal}(a), displays a low-lying branch analogous to that of the analytical Moore-Read state in Fig.~\ref{fig:MooreRead_BW_ideal}(a), as well as a novel high-lying branch~\cite{Sterdyniak_nJP_2011}. 
While the two branches are separated at low energy, they merge at higher energy, limiting the identification of the levels that belong to the low-energy branch.  
In the specific example of Fig.~\ref{fig:MooreRead_BW_nonIdeal}(a), we are able to unequivocally isolate the ground energy branch for 7 values of $L_z^A$, from $18$ to $24$.

We compare this entanglement spectrum with the energy spectrum of the corresponding BW Hamiltonian, shown in  Fig.~\ref{fig:MooreRead_BW_nonIdeal}(b). 
Just like the RSES, a single state emerges at angular momentum $L_z^A = 24$. As we move towards lower angular momentum states, we observe the development of a branch with the same number of states as the RSES, namely $\{1, 2, 4 \}$ from right to left. This branch eventually merges at angular momentum $L_z^A = 21$ with the high-lying branch (see Appendix~\ref{App:StateCounting:C} for more details on the degeneracy of these levels according to the arguments of conformal field theory). 

In order to investigate more thoroughly the connection between the BW and the real-space entanglement Hamiltonian,
we now consider bipartitions with particles numbers $N_A\neq N/2$.
A natural choice is to move the longitudinal bipartition, so that the probability of having $N_A$ particles within the spherical cap defined by $0\leq\theta\leq\theta_0$ (see the insets of Fig.~\ref{fig:SM_cuts}) does not change drastically as $N_A$ is varied (see Appendix~\ref{App:A} for a discussion on how to perform a cut that is not at the equator).

Such an angle can be determined by the following simple physical consideration.
A FQH ground state is an incompressible featureless liquid which in the thermodynamic limit has uniform density $\rho_0=\nu/(2\pi l_B^2)$, where $\nu$ is the filling fraction and $l_B=\sqrt{\hbar/qB}$ the magnetic length. We require
\begin{equation}
	N_A=\int_0^{2\pi}R\,d\phi\int_0^{\theta_0} R \sin(\theta)d\theta\,\rho_0
\end{equation}
i.e.~that the number of particles in the spherical-cap is precisely $N_A$. Since $R^2/l_B^2=Q$, this leads to $N_A=2\nu Q \sin^2\left(\frac{\theta_0}{2}\right)$. Since $2Q=N_\phi=\nu^{-1}N-\mathcal{S}\approx \nu^{-1}N$ ($\mathcal{S}$ being the so-called ``\textit{shift}"~\cite{WenZee_PRL_1992}) we finally get
\begin{equation}
	\label{eq:movingCut}
	\theta_0=2\arcsin\sqrt{\frac{N_A}{N}}.
\end{equation}
In the following, we will vary the cut longitude $\theta_0$ with $N$ and $N_A$ as prescribed by the relation in Eq.~\eqref{eq:movingCut}.
Notice that with this choice the length $\mathcal{L}$ of the cut does not change as the number of particles $N$ is increased and $N_A$ is kept constant, for $\mathcal{L}=2\pi R \sin(\theta_0)\approx \pi \sqrt{N_A/\nu}\, l_B$.

Comparisons between the real-space entanglement spectrum and the BW one are presented in Fig.~\ref{fig:SM_cuts} for $N=16$, $Q=7$ and three different values of $N_A=4,6,8$.
The location of the cuts at different values of $\theta_0$ can be inferred from the data in the insets, which display the ground-state density of the BW Hamiltonian.
The plots show that the counting of the lowest lying BW states matches that of the corresponding RSES.

These plots are an addition to
the data displayed in Fig.~\ref{fig:MooreRead_BW_nonIdeal} 
since they consider different values of $N_A$ with different parities.
As we discuss in Appendix~\ref{App:StateCounting:C}, for a Moore-Read state one expects a different degeneracy of the boundary levels depending on whether $N_A$ is even or odd. The values of $N_A$ considered in Fig.~\ref{fig:SM_cuts} are so small that finite size effects are important, and the degeneracy expected from conformal-field-theory arguments reported in Appendix~\ref{App:StateCounting:C} are only partially recovered; by considering larger setups one would indeed recover those numbers.

It can be noted moreover that
the lowest branch of the BW Hamiltonian spectrum merges with the upper one at lower values of $L_z^A$, covering a larger region.
Furthermore, the BW entanglement gap looks to be more robust for smaller $N_A$.
We do not have an explanation for this behaviour, which deserves further investigation.

We conclude this section with the statement that, even for this non-model wavefunction, the low-energy spectrum of the BW Hamiltonian reproduces the Li-Haldane conjecture.

\begin{figure}[t]
	\includegraphics[width=\columnwidth]{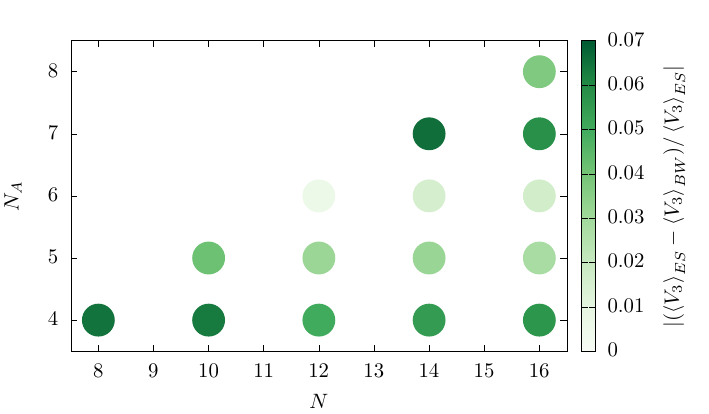}
	\caption{Absolute value of the relative error on the expectation value of the three-body contact interaction computed using either the eigenvectors of the real space entanglement Hamiltonian or the Bisognano-Wichmann one, in the case of a bosonic $\nu=1$ Moore-Read-like phase at $2Q=N-2$ stabilized by two-body contact interactions. 
		On the $x$-axis, various values for the number of particles $N$ (which fix the value of $Q$) are shown, while on the $y$-axis we plot various values for the particles $N_A$ in the northern spherical-cap.}
	\label{fig:SM_V3}
\end{figure}

\subsection{Eigenvectors}
We already discussed that in the case of model wavefunctions the eigenvectors of the BW Hamiltonian are the same as those of the real-space entanglement one. 
One does not expect this to be true in general for non-model states.
The question that we now address is to what extent the eigenvectors $\ket{\psi^{\rm BW}_i}$ of the BW Hamiltonian provide a good description of those of the entanglement one, $\ket{\phi^{\rm RSES}_\alpha}$.

We first consider the expectation value of the three-body contact interaction potential $V_3$, which is encoded in the colour of the markers of Figs.~\ref{fig:MooreRead_BW_nonIdeal} and~\ref{fig:SM_cuts}. 
This local observable quantifies the characteristic three-body anti-bunching expected in the Moore-Read state; the figure shows that it takes very similar values in both EH and BW Hamiltonians. 
In general, we observe a trend such that for small entanglement eigenvalues the expectation value $\langle V_3\rangle$ is small, and the latter grows by considering larger entanglement eigenvalues. This means that the density matrix $\rho_A$ is primarily composed of states where three particles are significantly anticorrelated. 

In order to be more quantitative, we focus on the ``entanglement ground-state", namely the state appearing with lowest entanglement eigenvalue.
In Fig.~\ref{fig:SM_V3} we compare the expectation values of the operator $V_3$ on the ``entanglement ground state'' of the exact EH and of the BW Hamiltonian, for several values of $\{N, N_A\}$. 
The relative error committed by the BW ansatz is within some percent points.

\begin{figure}[t]
	\includegraphics[width=\columnwidth,trim=0 1mm 0 0]{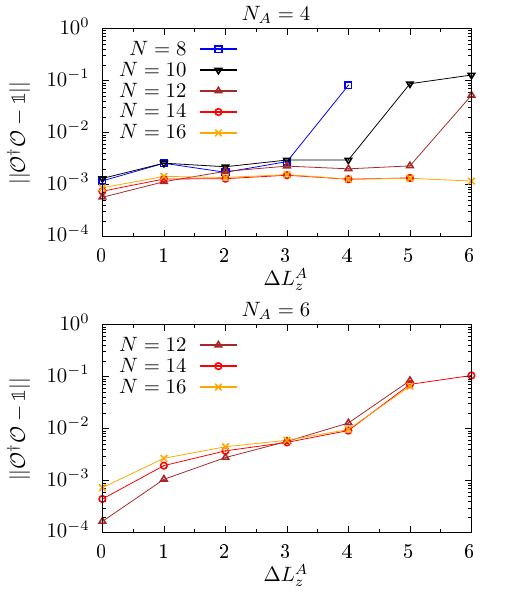}
	\caption{Deviations from unitarity of the matrix of overlaps between real-space entanglement and Bisognano-Wichmann Hamiltonian eigenvectors lying below the entanglement gap. Data are presented as a function of the angular-momentum variation $\Delta L_z^A$ with respect to the one of the entanglement Hamiltonian ground state, for different values of the number of particles $N$ (which fix the value of the flux through the sphere, $2Q=N-2$). The image on top considers $N_A=4$ particles in the northern spherical cap, the one at the bottom $N_A=6$.
	}
	\label{fig:SM_overlaps}
\end{figure}

Motivated by this result,
we compute the overlap matrix~\cite{Yan_PRB_2019} between the EH and BW Hamiltonian eigenvectors, within the low-energy branch, at fixed quantum numbers $N_A$ and $L_z^A$. We expect this matrix to describe a unitary change of basis when the BW Hamiltonian captures the EH physics. 
The results are presented in Fig.~\ref{fig:SM_overlaps} for $N_A=4$ and $N_A=6$, for different values of $L_z^A$ measured with respect to the ``entanglement ground-state" value. 
It can be seen that the BW Hamiltonian provides indeed a good description of the entanglement structure of fractional quantum Hall ground states, even away from the model-wavefunctions limit.
Summarizing, we find that $\mathcal O$ is a unitary matrix with a very high accuracy.	
We thus conclude that, for the values of $N$ and $N_A$ that we could consider, the EH and  BW Hamiltonians can be approximately put in direct correspondance in their low-energy branch.

We finally investigate the existence of any scaling limit making the BW Hamiltonian an exact expression of the EH; the data plotted in Figs.~\ref{fig:SM_V3} and~\ref{fig:SM_overlaps} do not make any simple trend apparent, and so far we could not find any. A possible reason could lie in the choice of the spherical geometry, which was chosen so as to avoid edge effects, but may be affected by Gaussian curvature effects.
The analysis of a planar geometry, which is free of curvature effects but may be plagued by edge ones, is reserved for a future study.
A second possibility could be that a system sufficiently large so that the long-wavelength regime in which the BW can be expected to give better results was not reached with the small sizes that can be studied with exact-diagonalization.
A hint of this fact can be seen even in the density plots in Fig.~\ref{fig:SM_cuts}, where density oscillations at the bipartition cut prevent the bulk from reaching the incompressible ``featureless" limit $\rho_0$.
Finally, a third possibility could be rooted in the fact that overlaps are not well defined in the thermodynamic limit, even though the proper long-wavelength features of the topological order have been captured. This would actually be analogous to what happens for model wavefunctions: their overlap with ``real" topologically ordered ground states is expected to drop to zero in the thermodynamic limit, even if the model wavefunction captures all of its essential details.
A better indicator could be to analyse expectation values of local observables, analogously to what was done in Fig.~\ref{fig:SM_V3}, but this again is reserved for future more systematic studies.

\section{Conclusions}\label{Sec:Conc}

We have revisited the fundamental problem of the RSES of FQH states proposing a study based on the BW Hamiltonian;
our results show that it correctly captures the features of the entanglement Hamiltonian, satisfying several properties among which the Li-Haldane conjecture.
These positive results open the exciting perspective of establishing an accurate mathematical connection beyond the numerical results that we have presented.

A remarkable result is that the BW Hamiltonian can be constructed straightforwardly from the knowledge of the physical Hamiltonian only, without the need of computing the ground state to be studied. 
This in particular means that one does not even need to know what specific form of topological order is displayed by the ground state, which is typically necessary for constructing the EH as an effective boundary conformal field theory. 
For this reason, we expect the BW Hamiltonian to become a useful tool in numerical studies of FQH systems.
The results presented in this article should motivate the use of the BW approach to study also other forms of topological order.

Finally, given the recent realisation of a non-interacting BW Hamiltonian in a cold-atom quantum simulator, this work opens new paths in the experimental study of the RSES of the paradigmatic topological phases of the FQH effects.

\textit{Acknowledgements.}---We acknowledge fruitful discussions with N. Regnault. We thank I.~Carusotto and J.~Dubail for careful reading of the manuscript.
This work is supported by European Union (grants TOPODY 756722 (Paris) and LoCoMacro 805252 (Orsay) from the European Research Council) and by Region Ile-de-France in the framework of the DIM Sirteq.
M. R. acknowledges support by the Deutsche Forschungsgemeinschaft (DFG) with project Grant No. 277101999 within the CRC network TR 183 (subproject B01) and under Germany's Excellence Strategy - Cluster of Excellence Matter and Light for Quantum Computing (ML4Q) EXC 2004/1 - 390534769.	
A. N. acknowledges support by \textit{Fondazione Angelo dalla Riccia} and thanks LPTMS for warm hospitality.
S. N. acknowledges support from Institut Universitaire de France.

\appendix

\vspace{-1mm}

\section{Details on the single-particle Bisognano-Wichmann Hamiltonian} \label{App:A}

In this Appendix we discuss the single-particle entanglement Hamiltonian and present the detailed calculations of the single-particle Bisognano-Wichmann Hamiltonian.

\subsection{Single-particle entanglement spectrum}
\label{App:A:EntHam}

We first analytically describe the structure of the entanglement Hamiltonian in the case of a single-particle state occupying a well defined orbital $Y_{Q,Q,m}$ in the lowest Landau level.
The reduced density matrix can be decomposed as
\begin{equation}
	\rho_A=\left(1-\mathcal{N}_A\right)\ket{v_A}\bra{v_A} + \mathcal{N}_A \ket{\psi_A}\bra{\psi_A} 
\end{equation}
where $A$ is the spherical cap defined by $\theta<\theta_0$, $\ket{v_A}$ is the state with no particles in $A$ and $\psi_A$ the one with a single particle restricted to this region. 
\begin{widetext}
Finally
\begin{equation}
	\label{eq:NA_sphere}
	\mathcal{N}_{A}=\int d\phi \int_0^{\theta_0} \sin(\theta)d\theta\, |Y_{Q,Q,m}|^2 = \frac{\mathrm{B}_{\sin^2\left(\theta_0/2\right)}\left(Q+1-m,Q+1+m\right)}{\mathrm{B}(Q+1-m,Q+1+m)}.
\end{equation}
Here, $\mathrm{B}(a,b)=\mathrm{B}_1(a,b)$ is the Beta function.
Since the particle either is the $A$ region or it is not, the entanglement Hamiltonian, defined by $\rho_A=e^{-H_A}/\text{Tr}[e^{-H_A}]$, can be written as
\begin{equation}
	H_A = E_{v_A} \ket{v_A}\bra{v_A} 
	+E_{\psi_A} \ket{\psi_A}\bra{\psi_A}
\end{equation}
with $E_{v_A}$ and $E_{\psi_A}$ being dimensionless quantities. 
With some algebra it can be shown that the previous relations are compatible provided
\begin{equation}
	\label{eq:NA_EE}
	\mathcal{N}_{A} = \frac{1}{2} \left[ 1 - \tanh \left(\frac{E_{\psi_A}-E_{v_A}}{2} \right) \right]
\end{equation}
which characterizes the entanglement energies up to a global additive constant, which can be fixed by requiring $E_{v_A}=0$.

Inverting Eq.~\eqref{eq:NA_EE} and using Eq.~\eqref{eq:NA_sphere} we get
\begin{equation}
	\label{eq:EntanglementSpectrum}
	\begin{split}
		E_{\psi_A} =& -2\,\text{atanh}\left(2\mathcal{N}_A-1\right)\\=&- 2\,\text{atanh}\left(2 \, \frac{\mathrm{B}_{\sin^2\left(\theta_0/2\right)}\left(Q+1-m,Q+1+m\right)}{\mathrm{B}(Q+1-m,Q+1+m)} - 1\right).
	\end{split}
\end{equation}
\end{widetext}
When $\theta_0=\pi/2$, Eq.~\eqref{eq:NA_sphere} can be linearized for $m/Q\ll1$ using the Beta function integral representation and Laplace's method, yielding
\begin{equation}
	\mathcal{N}_{A}\approx\frac{1}{2}+\frac{m}{\sqrt{\pi Q}}
\end{equation}
which, when inserted in Eq.~\eqref{eq:EntanglementSpectrum} gives
\begin{equation}
	E_{\psi_A} \approx - \frac{4m}{\sqrt{\pi Q}} = - \frac{4\,k l_B}{\sqrt{\pi}}.
\end{equation}
where the momentum $k=\frac{2\pi}{\mathcal{C}}\,m$ and $\mathcal{C}=2\pi R$ is the length of the bipartition cut. This last expression is the same one can find for an integer quantum Hall system on the plane~\cite{Oblak_PRB_2022}, which is locally indistinguishable from a sphere when its radius goes to infinity.

\subsection{Lowest band of the single-particle Bisognano-Wichmann  Hamiltonian}
In this subsection we show that the sharply-truncated lowest Landau level wavefunctions on the sphere are eigenvectors of the Bisognano-Wichmann  Hamiltonian.
In this supplemental materials, we discuss a general latitude angle $\theta_0$ for the bipartition, whereas in the main text we focused only on the half-bipartition of the sphere, $\theta_0=\pi/2$.

\begin{widetext}

We rewrite the Bisognano-Wichmann Hamiltonian as
\begin{equation}
	H_{BW} =\frac{4}{\sqrt{\pi\,Q}}\left\{
	\frac{H_0 - \epsilon}{\hbar\omega_c},
	\mathcal{F}(\theta)\right\}, 
	\qquad \text{with} \quad 
	0 \leq \theta \leq \theta_0 \quad \text{and} \quad
	\mathcal{F}(\theta)=Q(\cos(\theta)-\cos(\theta_0)) \geq 0; 
\end{equation}
the coordinates are restricted to the upper spherical cap and we introduced the function $\mathcal{F}(\theta)$ measuring the distance from the bipartition cut. 
In the fractional quantum Hall context, this distance function can interpreted as measuring the ``orbital-distance" between lowest-Landau level orbitals on the sphere, which are peaked at $\cos(\theta_M)=M/Q$. Notice that such a form appeared previously in Ref.~\cite{Bakas_AdS4_2015} within a different context.

Here $\epsilon$ is in principle an arbitrary energy shift, which we will choose by requiring the lowest Landau level truncated wavefunctions to be eigenfunction of the Bisognano-Wichmann Hamiltonian, 
and $H_0$ is the free single particle Hamiltonian \cite{jainCompositeFermionsBook_2007}
\begin{equation}
	H_0=\frac{|\mathbf{\Lambda}|^2}{2MR^2} =\frac{\hbar\omega_c}{2Q}\left(-\frac{1}{\sin(\theta)}\frac{\partial}{\partial\theta}\sin(\theta)\frac{\partial}{\partial\theta} + \left(Q\cot(\theta)+\frac{i}{\sin(\theta)}\frac{\partial}{\partial\phi}\right)^2\right).
\end{equation}
The (normalized) eigenvectors of $H_0$ are so-called monopole harmonics~\cite{jainCompositeFermionsBook_2007,wu_dirac_1976,Wu_PRD_1977}
\begin{equation}
	Y_{Q,l,m}=\sqrt{\frac{2l+1}{4\pi}\,\frac{(l-m)!}{(l-Q)!}\frac{(l+m)!}{(l+Q)!}}\,2^{-m}\,(1-\cos(\theta))^{\frac{-Q+m}{2}}(1+\cos(\theta))^{\frac{Q+m}{2}}\,P_{l-m}^{-Q+m,Q+m}(\cos(\theta))e^{i m \phi},
\end{equation}
where $P_n^{(\alpha,\beta)}(x)=2^{-n}\sum_{s=0}^{n}\binom{n+\alpha}{s}\binom{n+\beta}{n-s}(x-1)^{n-s}(x+1)^s$ are the Jacobi polynomials. 
Here, $2Q\in\mathbb{Z}$ is the flux through the sphere, $l\geq Q$ is the total angular momentum eigenvalue and $-l\leq m\leq l$ the eigenvalue of its projection along the $z$ axis.
The eigenvalue associated to $Y_{Q,l,m}$ is $E_{Q,l} = \hbar\omega_c \frac{l(l+1)-Q^2}{2Q}$; the lowest Landau level corresponds to $l=|Q|$, i.e. $E_{Q,Q} = \frac{1}{2}\hbar\omega_c$.

The action of $H_{BW}$ on these states can be computed
\begin{equation}
	\label{eq:BW_action}
	\frac{H_{BW}}{ 4/\sqrt{\pi Q} } Y_{Q,l,m} = \frac{1}{2Q}\left(
	-\frac{\partial^2\mathcal{F}}{\partial \theta^2} Y_{Q,l,m}
	-2\frac{\partial \mathcal{F}}{\partial \theta}\,\frac{\partial Y_{Q,l,m}}{\partial \theta} 
	-\frac{\cos(\theta)}{\sin(\theta)} \frac{\partial \mathcal{F}}{\partial \theta} Y_{Q,l,m} \right)  + 2\mathcal{F} \left(\frac{l(l+1)-Q^2}{2Q}-\frac{\epsilon}{\hbar\omega_c}\right) Y_{Q,l,m}.
\end{equation}
Notice that introducing Heaviside theta functions in the monopole harmonics (i.e. restricting them to the proper Bisognano-Wichmann domain) does not (as can be checked explicitly) change the above equation since the two domains $\theta>\theta_0$ and $\theta<\theta_0$ are disconnected by $\mathcal{F}(\theta)$, i.e. $\mathcal{F}(\theta_0)=0$.
Provided we set $\epsilon=\hbar\omega_c \frac{2Q+1}{2Q}$ (midway between the lowest LL $E_{Q,Q}$ and the first LL $E_{Q,Q+1}$), it is easy to check (to this purpose, see Eq.~\eqref{eq:monopole_derivative}) that $Y_{Q,Q,m}$ (the lowest-Landau level wavefunctions with $l=Q$) are eigenvectors
\begin{equation}
	\label{eq:lowestBWband}
	H_{BW} Y_{Q,Q,m} = -\frac{4}{\sqrt{\pi Q}}\left(m - (Q+1)\cos(\theta_0) \right)	Y_{Q,Q,m}.
\end{equation}
Analogously to lowest Landau level wavefunctions on the sphere, the normalized eigenvectors of the Bisognano-Wichmann Hamiltonian are most simply written in terms of spinor-variables $u(\theta, \phi)=\cos(\theta/2)e^{i\phi/2}$ and $v(\theta, \phi)=\sin(\theta/2)e^{-i\phi/2}$ as
\begin{equation}
	\label{eq:LBWL}
	\psi_{Q,m}^A = \frac{u^{Q+m}(\theta,\phi)v^{Q-m}(\theta,\phi)}{\sqrt{4\pi \mathrm{B}_{\sin^2\left(\frac{\theta_0}2\right)}(Q+1-m,Q+1+m)}}
	\Theta_{\rm H}(\theta_0-\theta),
\end{equation}
where $\mathrm{B}_x(a,b)$ is the incomplete Beta function, $\mathrm{B}_x(a,b)=\int_0^x t^{a-1}(1-t)^{b-1}\,dt$. 
Notice that we introduced the Heaviside theta function $\Theta_{\rm H}(\theta_0-\theta)$, which sharply separates the two spherical caps.

\end{widetext}

\subsection{Diagonalization of the single-particle Bisognano-Wichmann Hamiltonian}
In this subsection, we detail how the full spectrum of the Bisognano-Wichmann Hamiltonian is obtained.

We begin by introducing $\lambda_{Q,l}=l(l+1)-Q^2-(2Q+1)$ and rewriting Eq.~\eqref{eq:BW_action} as
\begin{align}
	\frac{H_{BW}}{ 4/\sqrt{\pi Q} } Y_{Q,l,m} =&(\lambda_{Q,l}+1)\,\cos(\theta) Y_{Q,l,m} + \nonumber \\
	&+\sin(\theta)\,\frac{\partial Y_{Q,l,m}}{\partial \theta} -  \lambda_{Q,l}\cos(\theta_0) Y_{Q,l,m}.
\end{align}

We now want to diagonalize $H_{BW}$ by expanding its general eigenvector on the basis provided by the monopole harmonics.
With the choice of bipartition $\mathcal{F}=\mathcal{F}(\theta)$ (independent of $\phi$), the $z$-projection of the angular momentum is still a good quantum number. We can therefore write
\begin{equation}
	H_{BW} \Psi_{Q,\alpha,m} = E_{Q,\alpha,m} \Psi_{Q,\alpha,m}
\end{equation}
and expand
\begin{equation}
	\Psi_{Q,\alpha,m} = \sum_{l} C_{l,\alpha}^{Q,m} Y_{Q,l,m}.
\end{equation}
multiplying by $Y^*_{Q,l',m}$ and integrating over the spherical cap $A$ we are considering we end up with a generalized eigenvalue problem (the states $Y_{Q,l,m}$ are not normalized nor orthogonal on the bipartition)
\begin{equation}
	\sum_{l} \mathcal{H}_{l',l}^{Q,m} C_{l,\alpha}^{Q,m} = E_{Q,\alpha,m} \sum_{l} \mathcal{M}_{l',l}^{Q,m} C_{l,\alpha}^{Q,m} 
	\label{Eq:Generalised:Eigenvalue}
\end{equation}
where
\begin{equation}
	\begin{cases}
		\mathcal{H}_{l',l}^{Q,m} = \int_{A} Y^*_{Q,l',m} H_{BW} Y_{Q,l,m} d\Omega
		\\
		\mathcal{M}_{l',l}^{Q,m} = \int_{A} Y^*_{Q,l',m} Y_{Q,l,m} d\Omega.
	\end{cases}
\end{equation}
The integrals could be performed numerically, but it is not difficult to express them as finite summations, which allow for a more straightforward numerical evaluation.
We here sketch the calculation of these matrix elements, giving only the main results.

The integration over the region $A$ can be seen as an integral over the whole space with a characteristic function $\Theta_{\rm H}(\theta_0-\theta)$ which can be expanded in terms of monopole harmonics
\begin{equation}
	\begin{cases}
		\Theta_{\rm H}(\theta_0-\theta) = \sum_l \zeta_l Y_{0,l,0}
		\\
		\zeta_l = 2\pi\sqrt{\frac{2l+1}{4\pi}}\,\frac{P_{l-1}(\cos(\theta_0)-P_{l+1}(\cos(\theta_0))}{2l+1}
	\end{cases}
\end{equation}
where $P_n(x)$ are Legendre polynomials.
\begin{widetext}
Products of monopole harmonics can be simplified in terms of Wigner 3j symbols~\cite{Wu_PRD_1977}
\begin{equation}
	\label{eq:monopole_harmonics_product_expansion}
	\begin{split}
		Y_{Q,l,m}Y_{Q',l',m'} =& \sum_{l''}(-1)^{l+l'+l''}(-1)^{Q+Q'}(-1)^{m+m'}\sqrt{\frac{(2l+1)(2l'+1)(2l''+1)}{4\pi}}\,\\&\threej{l}{l'}{l''}{m}{m'}{-(m+m')}\threej{l}{l'}{l''}{Q}{Q'}{-(Q+Q')}Y_{Q+Q',l'',m+m'}
	\end{split}
\end{equation}
and as a consequence of orthonormality integrals involving three of them can be written as~\cite{Wu_PRD_1977}
\begin{equation}
	\label{eq:triple_monopole_harmonics_integral}
	\int Y_{Q,l,m}Y_{Q',l',m'}Y_{Q'',l'',m''}d\Omega = (-1)^{l+l'+l''}\sqrt{\frac{(2l+1)(2l'+1)(2l''+1)}{4\pi}}\,\threej{l}{l'}{l''}{-m}{-m'}{-m''}\threej{l}{l'}{l''}{Q}{Q'}{Q''}.
\end{equation}
Using $Y^*_{Q,l,m}=(-1)^{Q-m}Y_{Q,l,m}$~\cite{Wu_PRD_1977}, the overlaps matrix elements can be written as
\begin{equation}
	\label{eq:overlaps_matrix}
	\mathcal{M}_{l',l}^{Q,m} = (-1)^{Q-m}\sum_{l''=|l'-l|}^{l+l'}\zeta_{l''}(-1)^{l+l'+l''}\sqrt{\frac{(2l+1)(2l'+1)(2l''+1)}{4\pi}}\,\threej{l}{l'}{l''}{-m}{m}{0}\threej{l}{l'}{l''}{Q}{-Q}{0}.
\end{equation}
Using $\cos(\theta) = \sqrt{\frac{4\pi}{3}}Y_{0,1,0}$ together with Eq.~\eqref{eq:monopole_harmonics_product_expansion} and Eq.~\eqref{eq:triple_monopole_harmonics_integral}, the first integral appearing inside $\mathcal{H}_{l,l'}^{Q,m}$ can be calculated
\begin{equation}
	\label{eq:first_contribution}
	\begin{split}
		I_{l',l}^{Q,m} = \int_{A} Y^*_{Q,l',m}\cos(\theta) Y_{Q,l,m} d\Omega=& 
		(-1)^{Q-m}\sum_{l'''=|l'-l|}^{l+l'}(2l'''+1) \threej{l}{l'}{l'''}{-m}{m}{0}\threej{l}{l'}{l'''}{Q}{-Q}{0}\\
		&\sum_{l''=|l'''-1|}^{l'''+1}(-1)^{1+l+l'+l''}\zeta_{l''} \sqrt{\frac{(2l+1)(2l'+1)(2l''+1)}{4\pi}}\threej{1}{l'''}{l''}{0}{0}{0}^2.
	\end{split}
\end{equation}
Using the results of Ref.~\cite{koornwinder_lowering_2005} it is not difficult to show that
\begin{equation}
	\label{eq:monopole_derivative}
	\sin(\theta)\,\frac{\partial Y_{Q,l,m}}{\partial \theta} = \left(l \cos(\theta)-\frac{Q}{l}m\right)Y_{Q,l,m}-\frac{1}{l}\sqrt{\frac{2l+1}{2l-1}(l^2-m^2)(l^2-Q^2)}\,\,Y_{Q,l-1,m}.
\end{equation}
The only contribution to $\mathcal{H}_{l,l'}^{Q,m}$ with a functional form different from the ones already analysed is seen to be
\begin{equation}
	\label{eq:second_contribution}
	\begin{split}
		J_{l',l}^{Q,m} = \int_{A} Y^*_{Q,l',m} Y_{Q,l-1,m} d\Omega=& 
		(-1)^{Q-m} \sum_{l''=|l-l'-1|}^{l+l'-1}\zeta_{l''}(-1)^{l-1+l'+l''}\sqrt{\frac{(2l-1)(2l'+1)(2l''+1)}{4\pi}}\,\\
		&\threej{l-1}{l'}{l''}{-m}{m}{0}\threej{l-1}{l'}{l''}{Q}{-Q}{0}.
	\end{split}
\end{equation}

Therefore, putting everything together we get
\begin{equation}
	\label{eq:final}
	\begin{split}
		\frac{\mathcal{H}_{l',l}^{Q,m}}{4/\sqrt{\pi Q}} =& \int_{A} Y^*_{Q,l',m} H_{BW} Y_{Q,l,m} d\Omega = (\lambda_{Q,l}+1)\,I_{l',l}^{Q,m}+\left(l\, I_{l',l}^{Q,m} - \frac{Q}{l} m \,\mathcal{M}_{l',l}^{Q,m} \right) -\\& \frac{1}{l}\sqrt{\frac{2l+1}{2l-1}(l^2-m^2)(l^2-Q^2)} \,\,J_{l',l}^{Q,m} - \lambda_{Q,l}\cos(\theta_0) \,\mathcal{M}_{l',l}^{Q,m}.
	\end{split}
\end{equation}
\end{widetext}
The matrix elements of both the Bisognano-Wichmann Hamiltonian $\mathcal{H}_{l',l}^{Q,m}$ and the overlaps matrix $\mathcal{M}_{l',l}^{Q,m}$ are henceforth obtained from Eq.~\eqref{eq:final}, Eq.~\eqref{eq:first_contribution}, Eq.~\eqref{eq:second_contribution} and Eq.~\eqref{eq:overlaps_matrix}; the generalized eigenvalue problem Eq.~\eqref{Eq:Generalised:Eigenvalue} can then be numerically solved.

Notice that the ground-band (see Eq.~\eqref{eq:lowestBWband}) can be  recovered in this formalism by setting $C_{l,\alpha_{gb}}^{Q,m}\propto \delta_{l,Q}$, meaning that $\Psi_{Q,\alpha_{gb},m}\propto Y_{Q,Q,m}$. 
This ansatz is indeed a valid solution for the generalized eigenvalue problem Eq.~\eqref{Eq:Generalised:Eigenvalue}
\begin{equation}
	\mathcal{H}_{l',Q}^{Q,m} = E_{Q,\alpha_{gb},m} \mathcal{M}_{l',Q}^{Q,m}
\end{equation}
provided $E_{Q,\alpha_{gb},m}$ are those given by Eq.~\eqref{eq:lowestBWband}.
Indeed after setting $l=Q$, the Hamiltonian matrix elements Eq.~\eqref{eq:final} simplify considerably
\begin{equation}
	\frac{\mathcal{H}_{l',Q}^{Q,m}}{4/\sqrt{\pi Q}} = -\left( m - (Q+1)\cos(\theta_0) \right) \,\mathcal{M}_{l',Q}^{Q,m}
\end{equation}
leading to
\begin{equation}
	E_{Q,\alpha_{gb},m} = - \frac{4}{\sqrt{\pi Q}} \left( m - (Q+1)\cos(\theta_0) \right)
\end{equation}
which as anticipated matches Eq.~\eqref{eq:lowestBWband}.

\section{Quantum numbers of the reduced density matrix}
\label{App:QuantumNumbersRedRho}

In this Appendix we briefly show that $N_A$ and $L^A_z$ are good quantum numbers of the reduced density matrix~\cite{regnault_entanglement_2017}.

Let us consider a quantum state defined on the sphere which has a well-defined number of particles $N$ and angular momentum $L_z$, dubbed $\ket{\Psi_{N, L_z}}$
By taking a bipartition at a given value of the polar angle $\theta$, that respects circular symmetry, it is possible to choose two bases for both regions $A$ and $B$ that have well-defined number of particles and angular momentum, dubbed $\{ \ket{N_A, L_z^A,i} \}$ and $\{ \ket{N_B, L_z^B,j} \}$, where $i$ and $j$ group all other quantum numbers. Hence, in principle the state $\ket{\Psi_{N, L_z}}$ should be expanded as:
\begin{widetext}
\begin{equation}
 \ket{\Psi_{N,L_z}} = \hspace{-0.25cm} \sum_{N_A, L_z^A, i, N_B, L_z^B,j} \hspace{-0.25cm}
 c_{N_A, L_z^A, i, N_B, L_z^B, j}
 \ket{N_A, L_z^A, i} \otimes \ket{N_B, L_z^B, j}.
\end{equation}
\end{widetext}
However, since the quantum numbers are constrained to satisfy the equalities $N_A+N_B = N$ and $L_z^A+L_z^B = L_z$, the expression can be simplified as:
\begin{widetext}
\begin{equation}
 \ket{\Psi_{N,L_z}} =  \sum_{N_A, L_z^A, i, j} 
 c_{N_A, L_z^A,i,j}
 \ket{N_A, L_z^A,i} \otimes \ket{N-N_A, L_z-L_z^A,j}.
\end{equation}
\end{widetext}

Let us now compute the reduced density matrix $\rho_A = \text{tr}_B[\ket{\Psi_{N,L_z}} \hspace{-0.1cm} \bra{\Psi_{N,L_z}}]$ using the basis $\{ \ket{N_B, L_z^B,j} \}$ introduced above:
\begin{widetext}
\begin{equation}
 \rho_A =  
 \sum_{N_A, L_z^A,i,i'} 
 \sum_j \left(
 c_{N_A, L_z^A,i,j} 
 c_{N_A, L_z^A,i',j}^* \right)
 \ket{N_A, L_z^A,i} \hspace{-0.1cm} \bra{N_A, L_z^A,i'}.
\end{equation}
\end{widetext}
We obtain an operator that in general does not have a well-defined value of $N_A$ or of $L_z^A$ but that is block-diagonal with respect to these quantum numbers.
In other words $\bra{N_A, L_z^A,i} \rho_A \ket{N_A', L_z^{A \prime}, i'}$ is different from zero only when $N_A=N_A'$ and $L_z^A = L_z^{A \prime}$ and equal to zero whenever $N_A \neq N_A'$ or $L_z^A \neq L_z^{A \prime}$.

As a consequence, the reduced density matrix $\rho_A$ commutes with $N_A$ and $L_z^A$ and so does also the entanglement Hamiltonian $H_A$; for this reason, the BW Hamiltonians that are proposed in this article commute with both observables.
In the plots that are presented in the main text, we focus on the eigenvectors of the entanglement or BW Hamiltonians for fixed values of $N_A$ and $L_z^A$.

\section{Diagonalization of the many-particles Bisognano-Wichmann Hamiltonian}\label{Appendix:Diag:ManyBody}
In this Appendix we briefly detail how the many-body problem has been diagonalized.
We consider, as we did in the main text, the two- and three- body contact interaction $V_2(\mathbf{r}_i,\mathbf{r}_j)=g_2 \delta^{(2)}(\mathbf{r}_i-\mathbf{r}_j)$ and 
$V_3(\mathbf{r}_i,\mathbf{r}_j,\mathbf{r}_k)=g_3 \delta^{(2)}(\mathbf{r}_i-\mathbf{r}_j)\delta^{(2)}(\mathbf{r}_i-\mathbf{r}_k)$. 
\begin{widetext}
The spatially deformed Bisognano-Wichmann interactions read
\begin{align}
	V_{A,2}^{\rm BW} &= \frac{8\sqrt{Q}}{\sqrt{\pi}}\,\frac{g_2}{R^2\hbar \omega_c}\,\frac{\delta(\theta_i-\theta_j)}{\sin(\theta_i)}\delta(\phi_i-\phi_j)\,\left(\cos(\theta_i)-\cos(\theta_0)\right)\\
	V_{A,3}^{\rm BW} &= \frac{8\sqrt{Q}}{\sqrt{\pi}}\,\frac{g_3}{R^4\hbar \omega_c}\,
	\frac{\delta(\theta_i-\theta_j)}{\sin(\theta_i)}\frac{\delta(\theta_i-\theta_k)}{\sin(\theta_i)}\delta(\phi_i-\phi_j)\delta(\phi_i-\phi_k)\,\left(\cos(\theta_i)-\cos(\theta_0)\right).
\end{align}
Since the magnetic field is assumed to be large, we project these deformed interactions onto the lowest level of the Bisognano-Wichmann Hamiltonian, Eq.~\eqref{eq:LBWL}. 
In second-quantization language the deformed interaction Hamiltonian therefore reads
\begin{align}
	\label{eq:BWProjectedInteractions_2}
	\Pi_{\rm BW}V_{A,2}^{\rm BW}\Pi_{\rm BW} &= \frac{1}{2!}\sum_{\substack{m_1,m_2\\m_3,m_4}} \bra{m_1,m_2}V_2\ket{m_3,m_4}\, a^\dagger_{m_1}a^\dagger_{m_2}a^\nodagger_{m_3}a^\nodagger_{m_4}\\
	\label{eq:BWProjectedInteractions_3}
	\Pi_{\rm BW}V_{A,3}^{\rm BW}\Pi_{\rm BW} &= \frac{1}{3!}\sum_{\substack{m_1,m_2,m_3\\m_4,m_5,m_6}} \bra{m_1,m_2,m_3}V_3\ket{m_4,m_5,m_6}\, a^\dagger_{m_1}a^\dagger_{m_2}a^\dagger_{m_3}a^\nodagger_{m_4}a^\nodagger_{m_5}a^\nodagger_{m_6}	
\end{align}
where $\ket{m}=a^\dagger_m \ket{0}$ and $a^\dagger_m$ creates a particle with angular momentum $m$ in the lowest-lying band of the Bisognano-Wichmann Hamiltonian; 
namely, $\left<\mathbf{r}|m\right> =\psi_{Q,m}^A(\mathbf{r})$ (see Eq.~\eqref{eq:LBWL}).
Since we consider bosons, annihilation/creation operators at the same angular momentum do not commute and rather obey $[a^\nodagger_{m},a^\dagger_{m'}]=\delta_{m,m'}$.

The relevant matrix elements can be computed explicitly as
\begin{align}
	\bra{m_1,m_2}V_2\ket{m_3,m_4} &=\frac{8\sqrt{Q}}{\sqrt{\pi}}\,\frac{g_2}{R^2\hbar \omega_c}\,\frac{\delta_{m_1+m_2,m_3+m_4}\,F_{\sin^2(\theta_0/2)}\left(Q,\frac{m_1+m_2+m_3+m_4}{2}\right)}{4\pi\sqrt{\prod_{i=1}^4 \mathrm{B}_{\sin^2(\theta_0/2)}\left(Q+1-m_i,Q+1+m_i\right)}}
	\\
	\bra{m_1,m_2,m_3}V_3\ket{m_4,m_5,m_6} &=\frac{8\sqrt{Q}}{\sqrt{\pi}}\,\frac{g_3}{R^4\hbar \omega_c}\,\frac{\delta_{m_1+m_2+m_3,m_4+m_5+m_6}\,G_{\sin^2(\theta_0/2)}\left(Q,\frac{m_1+m_2+m_3+m_4+m_5+m_6}{2}\right)}{(4\pi)^2\sqrt{\prod_{i=1}^6 \mathrm{B}_{\sin^2(\theta_0/2)}\left(Q+1-m_i,Q+1+m_i\right)}}
\end{align}
where
\begin{equation}
	\begin{split}
		F_x\left(Q,M\right) = \mathrm{B}_{x}\left(2Q-M+1,2Q+M+3\right) &-
		\mathrm{B}_{x}\left(2Q-M+3,2Q+M+1\right) -\\
		&-\cos(\theta_0)\,\mathrm{B}_{x}\left(2Q-M+1,2Q+M+1\right)
	\end{split}
\end{equation}
and
\begin{align}
	\begin{split}
		G_x\left(Q,M\right) = \mathrm{B}_{x}\left(3Q-M+1,3Q+M+3\right) &-
		\mathrm{B}_{x}\left(3Q-M+3,3Q+M+1\right) -\\
		&-\cos(\theta_0)\,\mathrm{B}_{x}\left(3Q-M+1,3Q+M+1\right).
	\end{split}
\end{align}
\end{widetext}
Using standard exact diagonalization techniques the projected Bisognano-Wichmann interaction Hamiltonians given by Eq.~\eqref{eq:BWProjectedInteractions_2} and Eq.~\eqref{eq:BWProjectedInteractions_3} can then be diagonalized. 

\section{State counting}
\label{App:StateCounting:C}

Before discussing the analysis we performed for the case of non-model wavefunctions, in this section we briefly review the counting of edge modes for the two topological orders we analysed, the Laughlin and the Moore-Read states.

\subsection{Laughlin edge-states counting}
A simple physical way of accounting for the edge excitations of a Laughlin state is the hydrodynamical approach.
This fractional quantum Hall state can be seen as a single-component incompressible fluid, which contains no low-energy bulk excitations. 
On the other hand, gapless chiral modes are localized at the system's boundary.
These surface deformations at low-energy and long-wavelengths can be completely described by a $\text U(1)$ Kac-Moody algebra~\cite{Wen_AdvPhys_1995}
\begin{equation}
	\arraycolsep=1.8pt\def\arraystretch{1.7}	
	\begin{array}{l}
		H=\frac{2\pi}{L}\hbar\frac{v}{\nu}\sum_{k>0} \rho_{-k}\,\rho_{k}\\
		\left[\rho_k,\rho_{-k'}\right]=\nu \frac{k L}{2\pi}\,\delta_{k,k'}	
	\end{array}
\end{equation}
where $\rho_k$ are the modes of the 1-dimensional density at the edge of the system.
We consider a compact edge of length $L$ and label the momentum $k=\frac{2\pi}{L}\,n$ by an integer $n$.
The edge of a Laughlin state can therefore be described by a bosonic field theory, with the Laughlin ground state being the vacuum of the theory $\ket{0}$.
The modes at a given momentum $K$ can then be written as $\prod_{n\in \boldsymbol{\lambda}} a^\dagger_n \ket{0}$, where $\boldsymbol{\lambda}$ is an integer partition of $K/(2\pi/L)\equiv M$ and $a^\dagger_n\propto \rho_{-k}$.
We list the lowest lying excitations in Table~\ref{tab:laughlinCounting}.

\begin{table}[t]
	\begin{center}
		\vspace{0.25cm}
		\begin{tabular}{c|c|c|c|c}
			\specialcell{Momentum\\ $\Delta M$} & \specialcell{Partition\\$\boldsymbol{\lambda}$} & States & \specialcell{Number\\ of states} & \specialcell{Energy\\ $\Delta E$}\\
			\hline
			\hline
			$0$ & $\{\}$ & $\ket{0}$ & 1 & $0$\\
			\hline
			$1$ & $\{1\}$ & $a^\dagger_1\ket{0}$ & 1 & $e_0$\\
			\hline
			$2$ & $\{2\}$ & $a_2^\dagger\ket{0}$ & 2 & $2e_0$\\			
			& $\{1,1\}$ & $(a^\dagger_1)^2\ket{0}$ & &\\
			\hline
			$3$ & $\{3\}$ & $a_3^\dagger\ket{0}$ & 3 & $3e_0$\\		
			& $\{2,1\}$ & $a_2^\dagger a_1^\dagger\ket{0}$ & &\\					
			& $\{1,1,1\}$ & $(a_1^\dagger)^3\ket{0}$ & &\\
			\hline
			$4$ & $\{4\}$ & $a_4^\dagger\ket{0}$ & 5& $4e_0$\\		
			& $\{3,1\}$ & $a_3^\dagger a_1^\dagger\ket{0}$& &\\					
			& $\{2,2\}$ & $(a_2^\dagger)^2\ket{0}$& &\\		
			& $\{2,1,1\}$ & $a_2^\dagger (a_1^\dagger)^2\ket{0}$& &\\
			& $\{1,1,1,1\}$ & $(a_1^\dagger)^4\ket{0}$& &\\
		\end{tabular}
		\vspace{0.25cm}
		\caption{\label{tab:laughlinCounting}
			List of the lowest lying edge excitations on top of the Laughlin ground state. We defined $e_0=v\,\hbar\frac{2\pi}{L}$. Momentum and energy are measured with respect to the one of the ground state, $\ket{0}$.}		
	\end{center}
\end{table}

\subsection{Moore-Read edge-states counting}
The edge excitations of the Moore-Read state contain, in addition to the $\text U(1)$ Kac-Moody algebra associated to a branch of bosonic edge excitations analogous to the one of the Laughlin liquid described above, a chiral Majorana fermion~\cite{Wen_PRL_1993,Milovanovic_PRB_1996}, which can have both periodic or anti-periodic boundary conditions. 
The low-energy and long-wavelength description of this mode reads
\begin{equation}
	\arraycolsep=1.8pt\def\arraystretch{1.7}	
	\begin{array}{l}
		H=\frac{2\pi}{L}\,\hbar v'\sum_{n\geq0} n\,c_{-n}\,c_{n}\\
		\left\{c_n,c_{-n'}\right\}=\delta_{n,n'},
	\end{array}
\end{equation}
where $n-\frac{1}{2}\in\mathbb{N}$ ($n\in\mathbb{N}$) for anti-periodic (periodic) boundary conditions on the fermion field.
The edge-excitations of the Moore-Read ground state turn out to correspond to the choice of anti-periodic boundary conditions. 
The theory has two additional independent sectors, according to the (conserved) parity of the number of fermions excited along the edge.
This parity has to match the one of the number of electrons in the system.
We list the lowest lying excitations in Table~\ref{tab:MajoranaFermionCounting_a} and \ref{tab:MajoranaFermionCounting_b}.

\begin{table}[b]
	\centering
			\begin{tabular}{c|c|c|c}
					\specialcell{Momentum\\ $\Delta M$} & States & \specialcell{Number\\of states} & \specialcell{Energy\\ $\Delta E$}\\
					\hline
					\hline
					$0$ & $\ket{0}$ & 1 & $0$\\
					\hline
					$1$ & & 0 & $e_0'$\\
					\hline
					$2$ & $c_{3/2} c_{1/2} \ket{0}$ & 1 & $2e_0'$\\
					\hline
					$3$ & $c_{5/2} c_{1/2}\ket{0}$ & 1 & $3e_0'$\\		
					\hline
					$4$ & $c_{7/2} c_{1/2}\ket{0}$ & 2& $4e_0'$\\		
					& $c_{5/2} c_{3/2}\ket{0}$& &\\
				\end{tabular}
				\caption{List of the lowest lying edge excitations of a Majorana fermion with antiperiodic boundary conditions and even number of fermions. 
					We defined $e_0'=\hbar v'\,\frac{2\pi}{L}$. Momentum and energy are measured with respect to the zero-momentum ground state, $\ket{0}$.}
				\label{tab:MajoranaFermionCounting_a}
\end{table}
\begin{table}
 			\centering
				\begin{tabular}{c|c|c|c}
					\specialcell{Momentum\\ $\Delta M$} & States & \specialcell{Number\\of states} & \specialcell{Energy\\ $\Delta E$}\\
					\hline
					\hline
					$0$ & $c_{1/2} \ket{0}$ & 1 & $0$\\					
					\hline
					$1$ & $c_{3/2} \ket{0}$ & 1 & $e_0'$\\
					\hline
					$2$ & $c_{5/2} \ket{0}$ & 1 & $2e_0'$\\
					\hline
					$3$ & $c_{7/2} \ket{0}$ & 1 & $3e_0'$\\		
					\hline
					$4$ & $c_{9/2} \ket{0}$ & 2& $4e_0'$\\		
					& $c_{5/2} c_{3/2} c_{1/2}\ket{0}$& &\\
				\end{tabular}
				\caption{List of the lowest lying edge excitations of a Majorana fermion with antiperiodic boundary conditions and odd number of fermions. 
					We defined $e_0'=\hbar v'\,\frac{2\pi}{L}$. Momentum and energy are measured with respect to the ground state, $c_{1/2} \ket{0}$.}
				\label{tab:MajoranaFermionCounting_b}
\end{table}

The total number of excitations of the Moore-Read state is then obtained by taking the product between the Majorana fermion sector and the bosonic charge sector. We list the number of degeneracies in Table~\ref{tab:MooreReadCounting_A} with a few selected examples for the states at $\Delta M=3$ in Table~\ref{tab:MooreReadCounting_B}.

\begin{table}
	\centering
				\begin{tabular}{c|c|c}
					\specialcell{Momentum\\ $\Delta M$} & \specialcell{Number of states \\ (even $N$)} & \specialcell{Number of states \\ (odd $N$)}\\
					\hline
					\hline
					$0$ & 1 & 1\\
					\hline
					$1$ & 1 & 2\\
					\hline
					$2$ & 3 & 4\\
					\hline
					$3$ & 5& 7\\
					\hline
					$4$ & 10& 13\\
				\end{tabular}
				\caption{\label{tab:MooreReadCounting_A} Number of gapless edge modes at the boundary of a Moore-Read topological order with either even or odd particle numbers, $N$.}	
\end{table}
\begin{table}
        \centering
				\vspace{0.25cm}
				\begin{tabular}{c|c|c}
					\specialcell{Momentum\\ $\Delta M$}&\specialcell{States \\ (even $N$)} & \specialcell{States \\ (odd $N$)}\\
					\hline
					\hline
					3&$a_3^\dagger\ket{0}$&$a_3^\dagger c^\nodagger_{1/2}\ket{0}$\\
					&$a_2^\dagger a_1^\dagger\ket{0}$&$a_2^\dagger a_1^\dagger c^\nodagger_{1/2}\ket{0}$\\
					&$(a_1^\dagger)^3\ket{0}$&$(a_1^\dagger)^3 c^\nodagger_{1/2}\ket{0}$\\
					&$a_1^\dagger c^\nodagger_{5/2}c^\nodagger_{1/2}\ket{0}$&$a_2^\dagger c^\nodagger_{3/2}\ket{0}$\\
					&$c_{5/2}^\nodagger c_{1/2}^\nodagger\ket{0}$&$(a_1^\dagger)^2 c^\nodagger_{3/2}\ket{0}$\\
					&&$a_1^\dagger c^\nodagger_{5/2}\ket{0}$\\
					&&$c^\nodagger_{7/2}\ket{0}$\\							
				\end{tabular}				
				\caption{\label{tab:MooreReadCounting_B} Edge-states of a Moore-Read topological order with either even or odd particle numbers, $N$ in the $\Delta M=3$ momentum sector.}
\end{table}

\newpage
	
\let\oldaddcontentsline\addcontentsline
\renewcommand{\addcontentsline}[3]{}
	

\providecommand{\noopsort}[1]{}

\end{document}